\documentclass[aps,pra,showkeys,twocolumn,superscriptaddress]{revtex4-2}
\usepackage{amsmath, amssymb}
\usepackage{mathrsfs}
\usepackage{array}
\usepackage{booktabs}
\usepackage{tabu}
\usepackage{dcolumn}
\usepackage{amsmath}
\usepackage{amsfonts}
\usepackage{float}
\usepackage{amssymb}
\usepackage{graphicx,color}
\usepackage[colorlinks={true}]
{hyperref}
\hypersetup{colorlinks=true,linkcolor=red,citecolor=blue,urlcolor=blue}
\usepackage{graphicx}
\usepackage{subfigure}
\usepackage{graphicx}
\usepackage{dcolumn}
\usepackage{bm}
\usepackage{pstricks}
\usepackage{braket}
\usepackage{orcidlink}

\def\be{\begin{equation}}
  \def\ee{\end{equation}}
\def\bea{\begin{eqnarray}}
\def\eea{\end{eqnarray}}
\def\f{\frac}
\def\n{\nonumber}
\def\l{\label}
\def\p{\phi}
\def\o{\over}
\def\R{\hat{\hat{\varrho}}}
\def\pa{\partial}
\def\om{\Omega}
\def\na{\nabla}
\def\P{\Phi}
\begin{document} 
\title{Kitaev Quantum Batteries: Super-Extensive Scaling of Ergotropy in 1D Spin$-1/2$ $XY-\Gamma(\gamma)$ Chain}

\author{Asad Ali\orcidlink{0000-0001-9243-417X}} 
\email{asal68826@hbku.edu.qa}
\affiliation{Qatar Centre for Quantum Computing, College of Science and Engineering, Hamad Bin Khalifa University, Doha, Qatar}

\author{Samira Elghaayda\orcidlink{0000-0002-6655-0465}}
\affiliation{Laboratory of High Energy Physics and Condensed Matter, Department of Physics, Faculty of Sciences of Aïn Chock, Hassan II University, Casablanca 20100, Morocco}

\author{Saif Al-Kuwari\orcidlink{0000-0002-4402-7710}}
\email{smalkuwari@hbku.edu.qa}
\author{M.I. Hussain\orcidlink{0000-0002-6231-7746}}
\email{mahussain@hbku.edu.qa}
\author{M.T. Rahim\orcidlink{0000-0003-1529-928X}}

\author{H. Kuniyil\orcidlink{0000-0003-0338-1278}} 
\affiliation{Qatar Centre for Quantum Computing, College of Science and Engineering, Hamad Bin Khalifa University,  Doha, Qatar}

\author{C. Seida\orcidlink{0000-0002-3842-3215}} 
\affiliation{Laboratory of R\&D in Engineering Sciences, Faculty of Sciences and
Techniques Al-Hoceima, Abdelmalek Essaadi University, Tetouan,
Morocco}

\author{A. El Allati\orcidlink{0000-0002-2465-8515}} 
\affiliation{Laboratory of R\&D in Engineering Sciences, Faculty of Sciences and
Techniques Al-Hoceima, Abdelmalek Essaadi University, Tetouan,
Morocco}
\affiliation{Max Planck Institute for the Physics of Complex Systems,
Nothnitzer Strasse 38, D-01187 Dresden, Germany}
\author{M. Mansour\orcidlink{0000-0003-0821-0582}} 
\affiliation{Laboratory of High Energy Physics and Condensed Matter, Department of Physics, Faculty of Sciences of Aïn Chock, Hassan II University, Casablanca 20100, Morocco}

\author{Saeed Haddadi\orcidlink{0000-0002-1596-0763}} 
\email{haddadi@semnan.ac.ir}
\affiliation{Faculty of Physics, Semnan University, P.O. Box 35195-363, Semnan, Iran}
\affiliation{Saeed's Quantum Information Group, P.O. Box 19395-0560, Tehran, Iran}
\date{\today}
\def\be{\begin{equation}}
  \def\ee{\end{equation}}
\def\bea{\begin{eqnarray}}
\def\eea{\end{eqnarray}}
\def\f{\frac}
\def\n{\nonumber}
\def\l{\label}
\def\p{\phi}
\def\o{\over}
\def\R{\hat{\hat{\varrho}}}
\def\pa{\partial}
\def\om{\Omega}
\def\na{\nabla}
\def\P{$\Phi$}
\begin{abstract}
We investigate the performance of a novel model based on a one-dimensional (1D) spin-$1/2$ Heisenberg $XY-\Gamma(\gamma)$ quantum chain, also known as 1D Kitaev chain, as a working medium for a quantum battery (QB) in both closed and open system scenarios. We analyze the closed QB scenario by analytically evaluating ergotropy across different spin-spin couplings, anisotropies in spin interactions, Zeeman field strengths, charging field intensities, $\Gamma$ interactions, and temperature. Our results indicate that the ergotropy is highly dependent on spin-spin coupling and anisotropy. 
Under variable parameters, an increase in the spin-spin coupling strength displays quenches and exhibits non-equilibrium trends in ergotropy. After a quench, ergotropy may experience a sharp increase or drop, suggesting optimal operational conditions for QB performance. 
In the open QB scenario, we examine spin chains of sizes $2 \leq N \leq 8$ under the influence of dephasing, focusing on the evolution of ergotropy. We study two charging schemes: parallel charging, where spins are non-interacting, and collective charging, involving spin-spin coupling.
In the former, increased Zeeman field strength enhances both the peak ergotropy and charging rate, although without any quantum advantage or super-extensive scaling. In the latter, increasing spin-spin coupling might not achieve super-extensive scaling without introducing anisotropy in the spin-spin interaction.
Our results suggest that optimal QB performance and a quantum advantage in scaling can be achieved by leveraging anisotropic spin-spin couplings and non-zero $\Gamma$ interactions, allowing for faster charging and higher ergotropy under super-extensive scaling conditions up to $\alpha=1.24$ for the given size of the spin chain. 

\end{abstract}
\maketitle
\section{Introduction}
Quantum Batteries (QBs) can leverage quantum coherence and correlations to offer advantages like enhanced energy density, faster charging, and super-extensive energy scaling \cite{alicki2013entanglement,campaioli2024colloquium,ferraro2018high,campaioli2018quantum,binder2015quantacell,shi2022entanglement,kamin2020entanglement,henao2018role,dou2022cavity,HaddadiQB2024}. By harnessing quantum correlations, QBs create new pathways for efficient energy storage \cite{shi2022entanglement,kamin2020entanglement,henao2018role,zhang2024quantum,ali2024ergotropy,ali2024magnetic,quach2023quantum,allahverdyan2004maximal,Gyhm2024}. However, the specific contributions of resources such as superposition, nonlocality, steering, entanglement, and discord to QB performance warrant deeper investigation \cite{shi2022entanglement,cruz2022quantum,zhang2024quantum,ali2024magnetic,song2024evaluating,ali2024ergotropy,zhang2024entanglement,cruz2022quantum}.
Using interacting many-body quantum systems, one can enhance QB performance, especially through collective effects that exceed those of non-interacting systems. Heisenberg spin chains, known for their rich quantum phases and behaviors, are well-suited as working mediums for QBs and are readily realizable across various platforms, including NMR systems \cite{buttgen2024magnetic,capponi2019nmr,van2024precise,ali2024magnetic}, trapped ions \cite{qiao2024tunable,kotibhaskar2024programmable,luo2024quantum,busnaina2024quantum}, Rydberg atoms \cite{kim2024realization,wang2024quantum}, and superconducting qubits \cite{rosenberg2024dynamics}.

Heisenberg spin chains exhibit unique non-local exchange interactions, such as Dzyaloshinsky-Moriya (DM) and Kaplan-Shekhtman-Entin-Wohlman-Aharony (KSEA) interactions beyond traditional ferromagnetic (FM) and antiferromagnetic  (AFM) spin configurations. To the best of our knowledge, current research on QBs with Heisenberg spin chains comprised of these interactions as working mediums has typically been limited to only two spins \cite{ali2024ergotropy,ali2024magnetic,zhang2024quantum}. Therefore, in this paper, we examine the $XY-\Gamma(\gamma)$ spin model, also called the 1D Kitaev model, with a focus on the $\Gamma(\gamma)$ interaction parameter. Here, $\gamma = -1$ corresponds to the DM interaction, $\gamma = +1$ to the KSEA interaction, and for any $\gamma$, this is generally called Kitaev-type interaction \cite{kheiri2024information,yang2020phase,zhao2022characterizing}. 
We analyze this model for QBs in both closed and open quantum system scenarios. For the closed QB, we use Jordan-Wigner, Fourier, and Bogoliubov transformations to diagonalize the $XY-\Gamma(\gamma)$ Hamiltonian under a transverse Zeeman field, identifying the thermal equilibrium state as the QB’s passive (uncharged) state. We then cyclically charge the QB using a NOT gate-based Hamiltonian. For the open QB, we assume the same passive state as the uncharged QB state, applying Pauli-$X$ noise and solving the Lindblad master equation to determine the QB’s state.

Our study has two main objectives: first, to optimize ergotropy in closed QB by examining various model parameters, for which we derive a closed-form expression for ergotropy applicable to any two spins on the chain. This enables us to study the maximum energy output over time under different conditions. Second, we analyze open QB by numerically solving the associated master equations under Pauli-$X$ noise. Here, we investigate how ergotropy scales with system size, showing that model parameters, interactions, and external fields can be fine-tuned to enhance energy extraction as the QB size increases. 

The paper is organized as follows: In Sec.~\ref{sec2}, we describe the one-dimensional spin-$1/2$ \(XY-\Gamma(\gamma)\) model, discuss the diagonalization process, and the algebra needed for the performance analysis of the QB in both closed and open quantum scenarios. Sec.~\ref{sec3} discusses the results of our simulations for closed and then open QB dynamics. Finally, Sec.~\ref{sec4} concludes this paper.

\section{Preliminaries}\label{sec2}

\subsection{QB Modeling}
A spin-based QB can be modeled using a quantum mechanical spin system characterized by specific coupling interactions. The Hamiltonian for nearest-neighbour $XY$-interaction for $N$-spin system can be written as

\begin{equation}
    \hat{\mathcal{H}}_{XY} = J \sum_{n=1}^{N-1} \left[\left(\frac{1+\delta}{2}\right) \hat{\sigma}_n^x \hat{\sigma}_{n+1}^x + \left(\frac{1-\delta}{2}\right) \hat{\sigma}_n^y \hat{\sigma}_{n+1}^y\right],
\end{equation}
where $\delta$ denotes the anisotropy parameter, governing the relative spin-spin interaction strengths along the $x$- and $y$-axes, and $\hat{\sigma}_n^{x(y)}$ is the $x(y)$-component of the Pauli matrix for the $n$-th spin. The coupling constant $J$ represents the strength of spin interactions in the $xy$-plane, with $J > 0$ indicating AFM coupling (opposite spin alignment) and $J < 0$ indicating FM coupling (parallel spin alignment). The magnitude and sign of $J$ critically influence the collective dynamics of the spin chain.

The $\Gamma$ interaction, which acts along the $z$-direction, can be modeled as:

\begin{equation}
    \hat{\mathcal{H}}_{\Gamma} = \Gamma \sum_{n=1}^{N-1} \left( \hat{\sigma}_n^x \hat{\sigma}_{n+1}^y + \gamma \hat{\sigma}_n^y \hat{\sigma}_{n+1}^x \right).
\end{equation}
Note that when $\gamma = -1$, $\Gamma$ corresponds to the $z$-component of DM interaction, whereas for $\gamma = +1$, it corresponds to the $z$-component of KSEA interaction \cite{ali2024study,ali2024ergotropy,zhang2024quantum} whereas for any $\gamma$, the $\Gamma$ corresponds to Kitaev interactions \cite{kheiri2024information}. 
The Hamiltonian for the working medium of the Kitaev quantum battery (KQB) is then given by:

\begin{equation}
\begin{aligned}
    \hat{\mathcal{H}}_{S} &= \hat{\mathcal{H}}_{XY} + \hat{\mathcal{H}}_{\Gamma} \\
    &= J \sum_{n=1}^{N-1} \left[\left(\frac{1+\delta}{2}\right) \hat{\sigma}_n^x \hat{\sigma}_{n+1}^x 
    + \left(\frac{1-\delta}{2}\right) \hat{\sigma}_n^y \hat{\sigma}_{n+1}^y \right] \\
    &\quad + \Gamma \sum_{n=1}^{N-1} \left( \hat{\sigma}_n^x \hat{\sigma}_{n+1}^y + \gamma \hat{\sigma}_n^y \hat{\sigma}_{n+1}^x \right).
\end{aligned}
\end{equation}

To guarantee the broken degeneracy and establish distinct energy levels for the QB \cite{le2018spin}, the Zeeman field along the $z$-direction is modeled by the following Hamiltonian:
\begin{equation}
    \hat{\mathcal{H}}_{F} = B \sum_{n=1}^N  \hat{\sigma}_n^z,
\end{equation}
where $B$ is the Zeeman splitting field strength and $\hat{\sigma}_n^z$ denotes the Pauli matrix along the $z$-axis for the $n$-th spin. With the constant Zeeman splitting Hamiltonian, one can finally get the complete Hamiltonian for our QB as  
\begin{equation}
\begin{aligned}
    \hat{\mathcal{H}}_{\mathcal{QB}} &= \hat{\mathcal{H}}_{S} + \hat{\mathcal{H}}_{F} \\
    &= J \sum_{n=1}^{N-1} \left[\left(\frac{1+\delta}{2}\right) \hat{\sigma}_n^x \hat{\sigma}_{n+1}^x 
    + \left(\frac{1-\delta}{2}\right) \hat{\sigma}_n^y \hat{\sigma}_{n+1}^y \right] \\
    &\quad + \Gamma \sum_{n=1}^{N-1} \left( \hat{\sigma}_n^x \hat{\sigma}_{n+1}^y + \gamma \hat{\sigma}_n^y \hat{\sigma}_{n+1}^x \right) 
    + B\sum_{n=1}^N \hat{\sigma}_n^z.
\end{aligned}
\label{eq:HQB}
\end{equation}

The diagonalization of this Hamiltonian with periodic boundary condition, i.e.,
$\hat{\sigma}_{N+1} = \hat{\sigma}_1$ is implemented in Appendix \ref{appendixA}. Therefore, the complete QB Hamiltonian after fermionization is given by
\begin{equation}
\begin{split}
\hat{\mathcal{H}}_{\mathcal{QB}} = \sum_{k>0}\hat{\mathcal{H}}_{k}= \sum_{k>0}& \big[(\mathcal{A}_k + P_k)\hat{c}_k^\dagger \hat{c}_k + (\mathcal{A}_k - P_k)\hat{c}_{-k}^\dagger \hat{c}_{-k} \\
&+(Q_k + i\mathcal{B}_k)\hat{c}_k^\dagger \hat{c}_{-k}^\dagger \\
&+ (-Q_k + i\mathcal{B}_k)\hat{c}_k\hat{c}_{-k} - 2B\big],
\end{split}
\end{equation}
where $\mathcal{A}_k = 2[J\cos(k) + B]$, $\mathcal{B}_k = 2J\delta\sin(k)$, $P_k = 2\Gamma (\gamma - 1) \sin(k)$, and $Q_k = 2\Gamma (\gamma + 1) \sin(k)$.

\subsection{QB Charging}
Before charging the QB, it can generally be considered to be in its ground state \cite{zhang2023quantum,joshi2022experimental,song2024evaluating,lai2024quick} corresponding to absolute zero or in a thermal state at a finite temperature \cite{ghosh2020enhancement,quach2023quantum}. In the earlier case, the QB starts in its ground state and transitions to an excited state through a unitary operator, charging the battery over time. We adopt the thermal state option because operating a QB at absolute zero is not feasible in practice. For any general Hamiltonian $\hat{\mathcal{H}}_{\mathcal{}}$, one can write a general form of the Gibbs state \cite{binder2015quantum}, representing a system in thermal equilibrium at temperature $T$, is given by
\begin{equation}
 \hat{\zeta} = \mathcal{Z}^{-1} e^{-\beta \hat{\mathcal{H}}_{\mathcal{}}},  
\end{equation}
where $ \beta = 1/k_B T$ is the inverse temperature (set $k_B=1$) and $\mathcal{Z} = \text{Tr}(e^{-\beta \hat{\mathcal{H}}_{\mathcal{}}})$ is the partition function.

The QB can be charged by applying a local external magnetic field in either the $x$-direction or the $y$-direction, with the charging field strength denoted by $\Omega(t)$. 
We observe that the Pauli $\hat{\sigma}^x$ and $\hat{\sigma}^y$ gates are effective for charging a QB because they induce transitions between the quantum states, enabling population inversion. This allows the working medium to absorb energy from an external field and transition between different states, which is essential for the charging process. In contrast, the Pauli $\hat{\sigma}^z$ gate only shifts the energy levels and thus increases the energy scale without causing transitions between the states, making it ineffective for population inversion. Although $\hat{\sigma}^z$ can create an energy gap and remove degeneracy, it does not facilitate the energy exchange required to charge the QB, which is why the Zeeman splitting field is applied along the $z$ direction.

We assume that the charger provides constant charging in time; therefore, the charging field strength is a constant function of time, $\Omega(t)=\Omega$. The corresponding local charging Hamiltonian $\hat{\mathcal{H}}_{\mathcal{C}}$, is given by 
\begin{equation}
    \hat{\mathcal{H}}_{\mathcal{C}}^{(x,y)} = \Omega \sum_{n=1}^{N} \hat{\sigma}_n^{(x,y)}.
    \label{EQ8}
\end{equation}

\begin{figure*}[t]
  \centering  
  \includegraphics[width=0.4\textwidth]{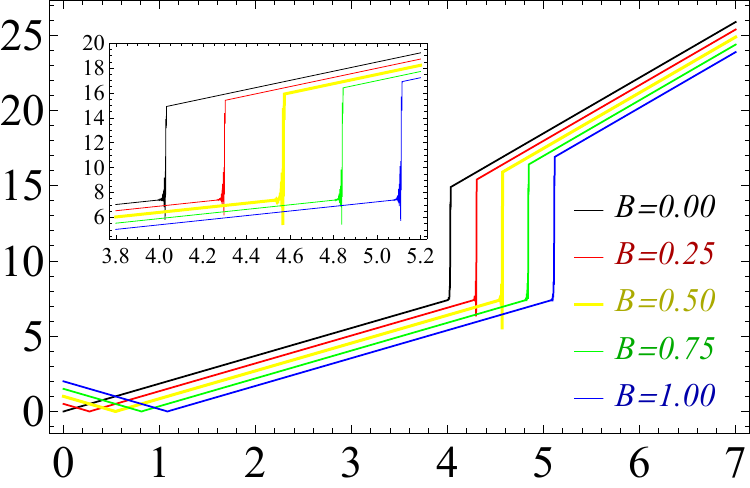}
  \put(-215,60){\rotatebox{90}{$\xi_{\max}^c$}}
  \put(-170,45){(a)}
  \put(-100,25){$T=0.01$} 
  \put(-97,-8){$J$} 
  \qquad  \qquad 
 \includegraphics[width=0.39\textwidth]{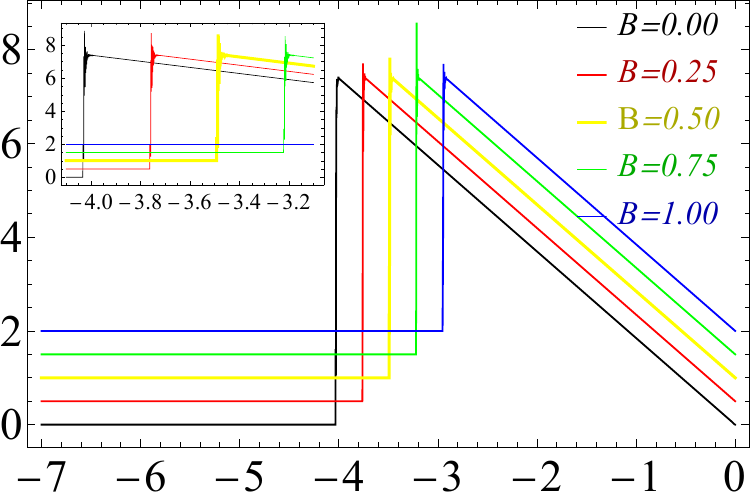}
  \put(-215,60){\rotatebox{90}{$\xi_{\max}^c$}}
  \put(-170,55){(b)}
  \put(-95,25){$T=0.01$} 
  \put(-97,-8){$J$} 
  \qquad  
\includegraphics[width=0.405\textwidth]{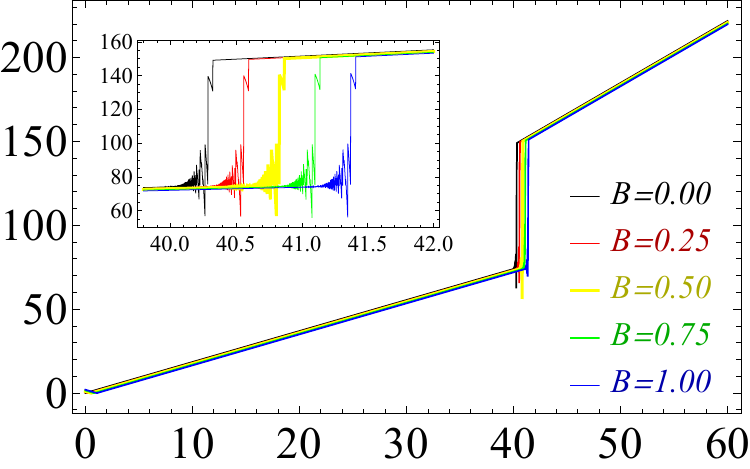}
  \put(-215,60){\rotatebox{90}{$\xi_{\max}^c$}}
  \put(-170,45){(c)}
  \put(-100,25){$T=0.1$} 
  \put(-97,-8){$J$} 
  \qquad \qquad  
  \includegraphics[width=0.4\textwidth]{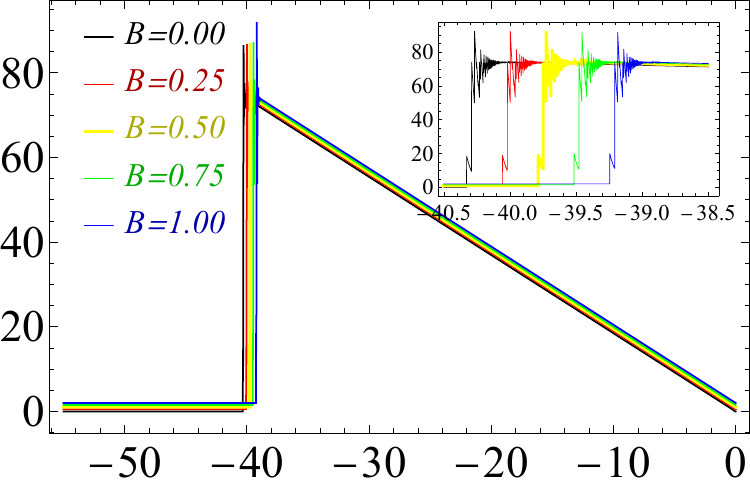}  
  \put(-215,60){\rotatebox{90}{$\xi_{\max}^c$}}
  \put(-170,50){(d)}
  \put(-95,25){$T=0.1$} 
  \put(-97,-8){$J$} 
\caption{Variation of maximum value of ergotropy in time, $\xi_{\max}^c$, as a function of $J$ for different fixed values of Zeeman splitting field such as $B=0$ (black), $B=0.25$ (red), $B=0.5$ (yellow), $B=0.75$ (green), and $B=1$ (blue) at $\Gamma=\gamma=0$. In (a-b), we have $T=0.01$, whereas in (c-d), we set $T=0.1$. Furthermore $\delta=0$, $k=7\pi/8$, and $\Omega=1$. Graphs (a) and (c) correspond to the AFM case whereas (b) and (d) correspond to the FM scenario.}
\label{F1}
\end{figure*}
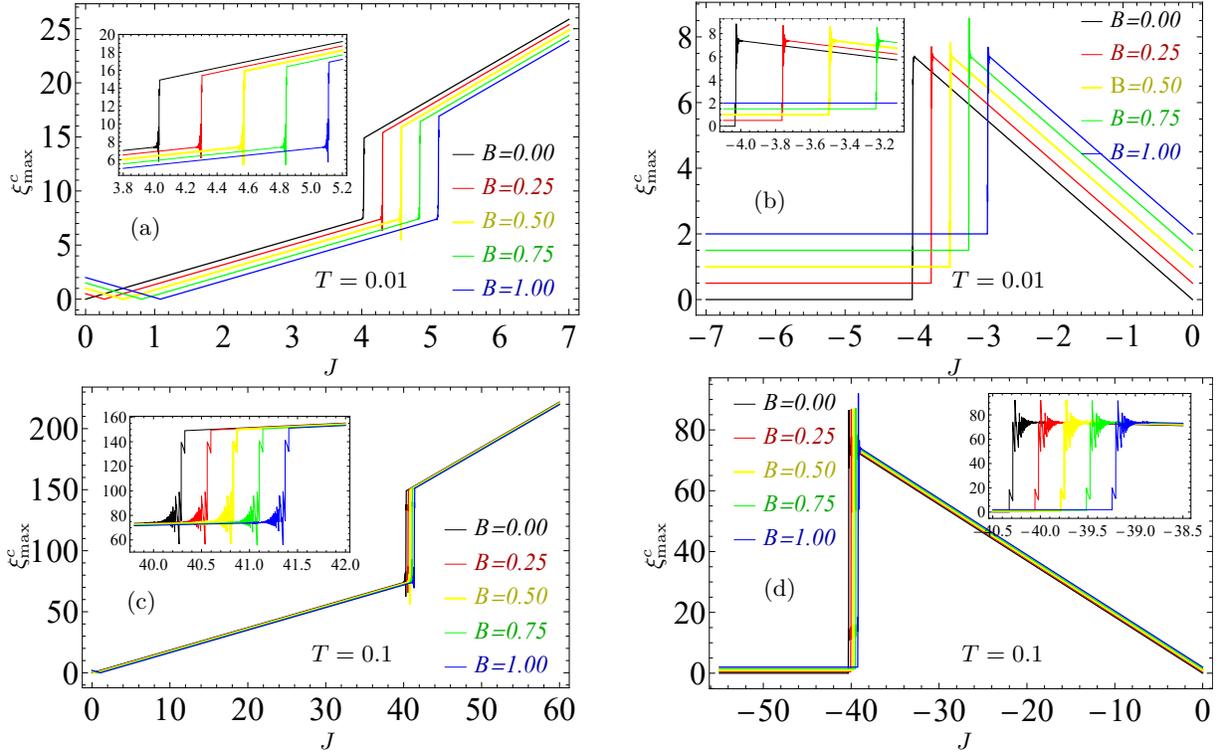

We will only consider the Pauli-X gate charging Hamiltonian henceforth. That is, $\hat{\mathcal{H}}_{\mathcal{C}}^{(x)}$ \cite{ghosh2021fast,ghosh2020enhancement,ali2024ergotropy}.

\subsection{Instantaneous state}
During the charging process, the state of closed QB $\hat{\varrho}_{c} (t)$ can be evaluated via
\begin{equation}
\hat{\varrho}_{c}(t)= \hat{U}_{\mathcal{C}}^{(x)}(t) \hat{\zeta}_{c} \hat{U}_{\mathcal{C}}^{(x)\dagger} (t),
\label{EQ9}
\end{equation}
where $\hat{\zeta}_{c}$ is uncharged Gibbs state of the closed QB, and 
\begin{equation}
\hat{U}_{\mathcal{C}}^{(x)}(t)=\exp(-i \hat{\mathcal{H}}_{\mathcal{C}}^{(x)}t),
\label{EQ10}
\end{equation}
is a charging unitary operator \cite{ghosh2021fast,ghosh2020enhancement,ali2024ergotropy}.

Open quantum system scenarios accommodate for dissipation by linking the system to an environment, leading to the evolution of the state $\hat{\varrho}_{o}(t)$ of open QB governed by the Lindblad master equation \cite{zhao2021quantum,yao2022optimal,shastri2024dephasing,ahmadi2024nonreciprocal,rodriguez2023catalysis}
\begin{align}\label{eq: pauli_lind}
\frac{d\hat{\varrho}_{o}(t)}{dt} &= -i\left[\hat{\mathcal{H}}_{\mathcal{QB}} + \hat{\mathcal{H}}_{\mathcal{C}}^{(x)}, \hat{\varrho}_{o}(t)\right] \nonumber \\
&\quad + \sum_k \left( L_k \hat{\varrho}_{o}(t) L_k^\dagger 
- \frac{1}{2} \left\{ L_k^\dagger L_k, \hat{\varrho}_{o}(t) \right\} \right).
\end{align}
We consider Pauli-$X$ dephasing noise, described by the Lindblad operator $L_k = \sqrt{g}\sigma_i^x$, which induces dephasing effects that influence the coherence of the quantum state. At $t=0$, the QB is initially in a thermal state $\hat{\zeta}_o$.

\subsection{QB Work Extraction}\label{subsec3B}
The energy extracted in the form of work from the QB can generally be written as \cite{zhao2021quantum,quach2023quantum,ali2024ergotropy} 
\begin{equation}
\mathcal{W}^{c(o)}(t)=\text{Tr}\left[(\hat{\varrho}_{c(o)}(t) -\hat{\varrho}(0)_{c(o)})\mathcal{\hat{H}}_{\mathcal{QB}}\right].
\end{equation}
This work definition is general and applies to closed and open QB settings. Now, if the uncharged state of QB is passive, then the work done is equivalent to ergotropy, which is defined as the maximum amount of cyclic work extracted from the QB \cite{binder2015quantum,sparaciari2017energetic,ali2024ergotropy}. In our case, the uncharged state is Gibbs state in both closed and open charging scenarios; therefore, we define the ergotropy as 

\begin{equation}
\xi^{c(o)}(t)=\text{Tr}\left[(\hat{\varrho}_{c(o)}(t) -\hat{\zeta}_{c(o)})\mathcal{\hat{H}}_{\mathcal{QB}}\right].
\label{EQ13}
\end{equation}

We use the ergotropy as a figure of merit to analyze the performance of both closed and open QB cases.
\section{Results and Discussions}\label{sec3}
In this section, we discuss both the closed and open KQB scenarios. We begin by investigating the performance of closed QB, focusing on two spin sites on the $XY-\Gamma$ chain. We analyze the ergotropy against various parameters of the working medium of the QB, including the applied Zeeman field strength and charging field strength. In the open QB scenario, we employ a numerical approach rather than an analytical one due to the complexity introduced by multiple spins. This approach is essential for examining the extensive scaling of output energy, which requires increasing the number of spins to effectively manifest the super-extensive scaling behavior in order to identify the quantum advantage and acquire the appropriate estimates of super-extensive scaling exponents.

\subsection{Closed QB scenario}
To analyze the performance of closed QB, we take two spin sites $k$ and $-k$ from the chain to deal with it analytically \cite{kheiri2024information}. We can rewrite the Hamiltonian $\hat{\mathcal{H}}_{\mathcal{QB}} = \sum_{k>0} \hat{\mathcal{H}}_k$ in the eigenbasis $\{|0_k0_{-k}\rangle, |1_k1_{-k}\rangle, |1_k0_{-k}\rangle, |0_k1_{-k}\rangle\}$ in the form 

\begin{align}
\hat{\mathcal{H}}_k = & -2B \bigg( |0_k0_{-k}\rangle \langle 0_k0_{-k}| 
+ |0_k 1_{-k}\rangle \langle 0_k 1_{-k}| \bigg) \nonumber \\
& -2B \bigg( |1_k 0_{-k}\rangle \langle 1_k 0_{-k}| 
+ |1_k 1_{-k}\rangle \langle 1_k 1_{-k}| \bigg) \nonumber \\
& + (Q_k - i\mathcal{B}_k) |0_k 0_{-k}\rangle \langle 0_k 1_{-k}| \nonumber \\
& + (Q_k + i\mathcal{B}_k) |0_k 1_{-k}\rangle \langle 0_k 0_{-k}| \nonumber \\
& + (-2B + 2\mathcal{A}_k) |0_k 1_{-k}\rangle \langle 0_k 1_{-k}| \nonumber \\
& + (-2B + \mathcal{A}_k + P_k) |1_k 0_{-k}\rangle \langle 1_k 0_{-k}| \nonumber \\
& + (-2B + \mathcal{A}_k - P_k) |1_k 1_{-k}\rangle \langle 1_k 1_{-k}|.
\label{EQ14}
\end{align}

As mentioned before, we assume that the initial state of the QB, $\hat{\zeta}_{c}$, is a Gibbs thermal state which reads  
\begin{equation}
\hat{\zeta}_{c} =\mathcal{Z}_c^{-1} \text{Tr}(e^{-\beta \hat{\mathcal{H}}_{k}})=\frac{1}{\mathcal{Z}_c}
\begin{pmatrix}
\zeta_{11} & \zeta_{12} & 0 & 0 \\
\zeta_{21} & \zeta_{22} & 0 & 0 \\
0 & 0 & \zeta_{33} & 0 \\
0 & 0 & 0 & \zeta_{44}
\end{pmatrix},
\label{EQ15}
\end{equation}
where $\mathcal{Z}_c = 2\left[\cosh(\beta\Lambda_k) + \cosh(\beta P_k)\right]$ is the partition function and
\begin{align}
\zeta_{11/22} &= \cosh(\beta\Lambda_k) \pm \cos(2\Phi_k)\sinh(\beta\Lambda_k), \nonumber \\
\zeta_{12/21} &= -e^{\mp i\theta_k}\sin(2\Phi_k)\sinh(\beta\Lambda_k), \nonumber \\
\zeta_{33/44} &= e^{\pm\beta P_k},
\label{eq:d_elements}
\end{align}
here, one can define $\Lambda_k$ as:
\begin{equation}
\Lambda_k = \sqrt{\mathcal{A}_k^2 + \mathcal{B}_k^2 + Q_k^2}
\end{equation}
and $\Phi_k$ as:
\begin{equation}
\Phi_k = \frac{1}{2} \arctan \left( \operatorname{sgn}(k) \frac{\sqrt{\mathcal{B}_k^2 + Q_k^2}}{\mathcal{A}_k} \right),
\end{equation}
where $\operatorname{sgn}(k)$ denotes the sign function, such that $\operatorname{sgn}(k) = 1$ for $k \geq 0$ and $\operatorname{sgn}(k) = -1$ for $k < 0$.

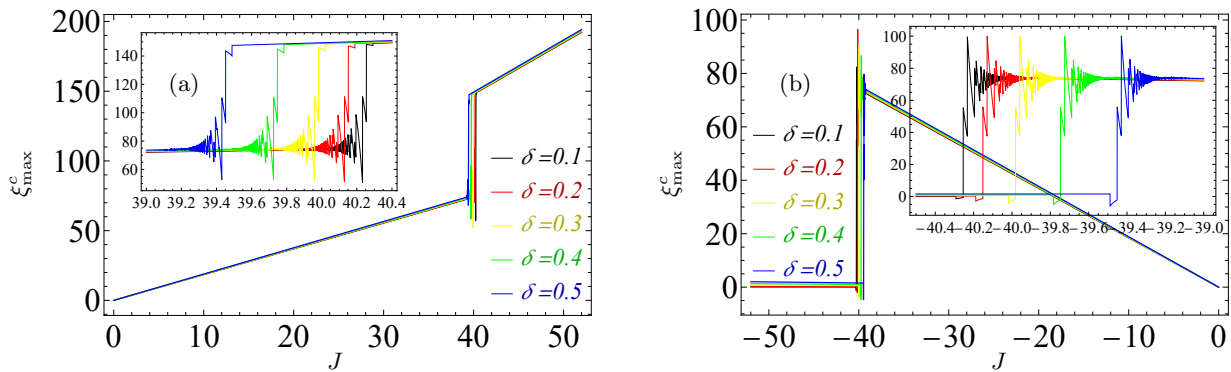
\begin{figure*}[t]
  \centering  
  \includegraphics[width=0.4\textwidth]{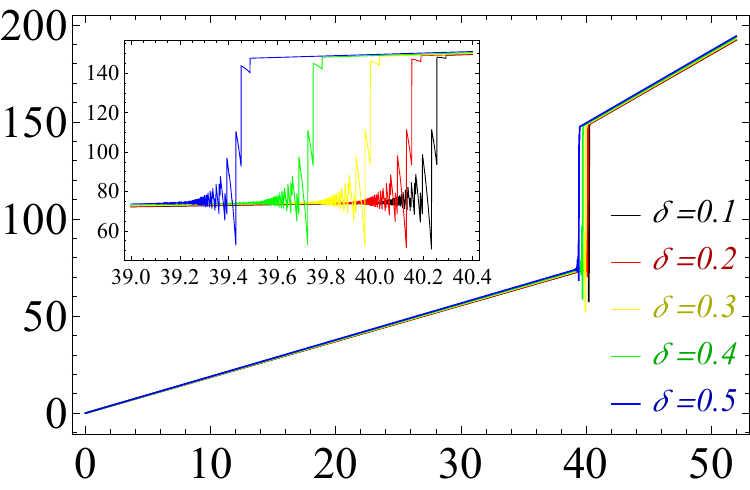}
  \put(-220,60){\rotatebox{90}{$\xi_{\max}^{c}$}}
  \put(-160,100){(a)} 
  \put(-100,-5){$J$} 
  \put(-110,20){$$} 
  \qquad \qquad  
  \includegraphics[width=0.4\textwidth]{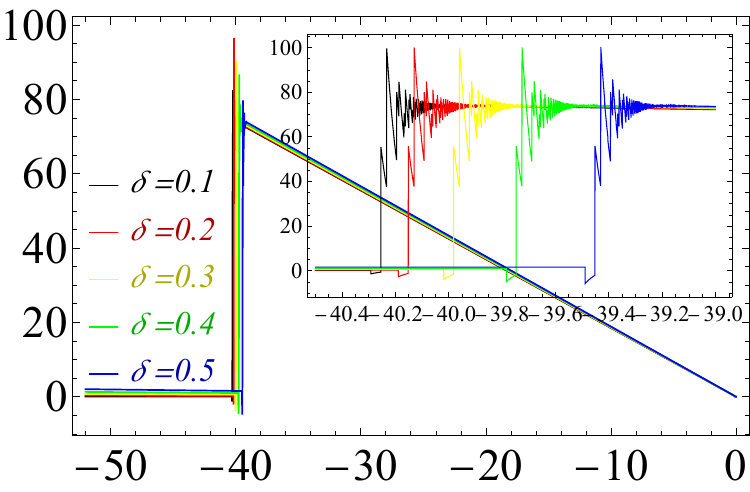}
  \put(-215,60){\rotatebox{90}{$\xi_{\max}^{c}$}}
  \put(-170,100){(b)} 
  \put(-90,-5){$J$} 
  \put(-110,20){$$} 
  \qquad
\caption{Evolution of maximum value of ergotropy in time, $\xi_{\max}$, as a function of $J$ for different fixed values of $\delta=0.1$ (black), $\delta=0.2$ (red), $\delta=0.3$ (yellow), $\delta=0.4$ (green), and $\delta=0.5$ (blue) at $\Gamma=B=\gamma=0$, $T=0.1$, $k=7\pi/8$, and $\Omega=1$. Here (a) corresponds to AFM whereas (b) is the FM case except at $J=0$.}
\label{F2}
\end{figure*}
In the closed QB case, the charging process can be implemented as a unitary evolution for the charger's Hamiltonian. From Eq. \eqref{EQ8}, the Hamiltonian for any two spins based on Pauli-$X$ gate charging operator takes the following form: 

\begin{equation}
\hat{\mathcal{H}}_{\mathcal{C}~(k,-k)}^{(x)} = \Omega \left(\hat{\sigma}_{(k)}^{(x)} \otimes \hat{\mathbb{I}}_{(-k)} + \hat{\mathbb{I}}_{(k)} \otimes \hat{\sigma}_{(-k)}^{(x)}\right).
    \label{EQ19}
\end{equation}

Therefore, substituting Eq. \eqref{EQ19} in $\hat{U}_{\mathcal{C}}^{(x)}(t)=\exp(-i\hat{\mathcal{H}}_{\mathcal{C}~(k,-k)}^{(x)}t)$, the charging unitary takes the following form: 

\begin{equation}
     \hat{U}_{\mathcal{C}}^{(x)} (t) = \left(
\begin{array}{cccc}
 a & c & c & b \\
 c & a & b & c \\
 c & b & a & c \\
 b & c & c & a \\
\end{array}
\right),
\end{equation}
where $a = \cos^2(\Omega t)$, $b = -\sin^2(\Omega t)$, and $c = -\frac{i}{2} \sin(2 \Omega t)$.

Using the Gibbs state from Eq.~\eqref{EQ15} as the initial state and the charging operator presented in Eq.~\eqref{EQ19}, the instantaneous state of the closed QB during the cyclic unitary charging process can be determined. Subsequently, the ergotropy can be analytically computed using Eq.~\eqref{EQ13}. Although this ergotropy expression serves as the basis for the subsequent analysis of the closed QB, its complete analytical form is omitted here due to its complexity and length.

\begin{figure}[t]
    \centering
    \includegraphics[width=0.85\linewidth]{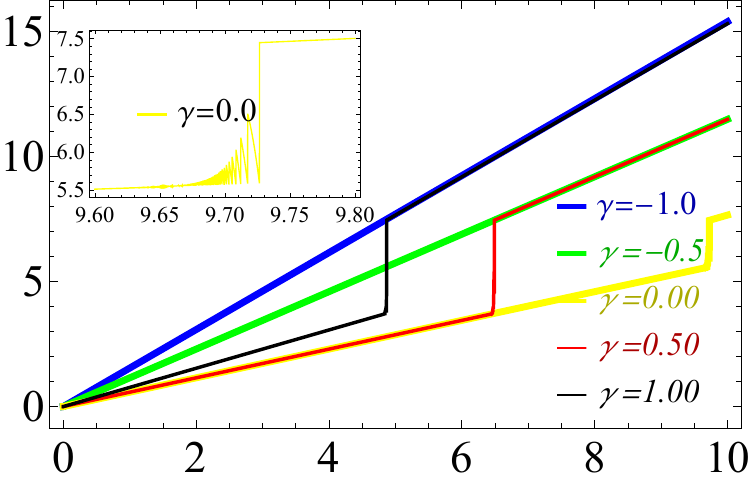}
    \put(-210,60){\rotatebox{90}{$\xi_{\max}^{c}$}}
   \put(-95,-5){$\Gamma$} 
      \caption{Evolution of maximum value of ergotropy in time, $\xi_{\max}^{c}$, as a function of $\Gamma$ for different fixed values of $\gamma=1.0$ (black), $\gamma=0.5$ (red), $\gamma=0.0$ (yellow), $\gamma=-0.5$ (green), and $\gamma=-1.0$ (blue) at $J=\delta=B=0$, $T=0.01$, $k=7\pi/8$, and $\Omega=1$. }
    \label{F3}
\end{figure}

Fig. \ref{F1} presents the maximum ergotropy over time, denoted as $\xi^{c}_{\max}$, plotted as a function of the spin-spin coupling on $xy$-plane $J$, for several fixed values of the Zeeman splitting field: $B = 0.00$ (black curve), $B = 0.25$ (red curve), $B = 0.50$ (yellow curve), $B = 0.75$ (green curve), and $B = 1.00$ (blue curve). We observe that the maximum charging time to take the Gibbs state as an uncharged state to a charged state is $\frac{\pi}{4\Omega}$. Panels \ref{F1}(a) and \ref{F1}(c) correspond to temperatures $T = 0.01$ and $T = 0.1$, respectively, focusing on positive values of $J$, which describe the AFM spin configuration. Conversely, panels \ref{F1}(b) and \ref{F1}(d), also at $T = 0.01$ and $T = 0.1$, respectively, highlight the behavior for negative values of $J$, representing the FM spin configuration. In Fig. \ref{F1}, the other parameter values are $\Gamma = 0$, $\gamma = 0$, and $\delta = 0$.

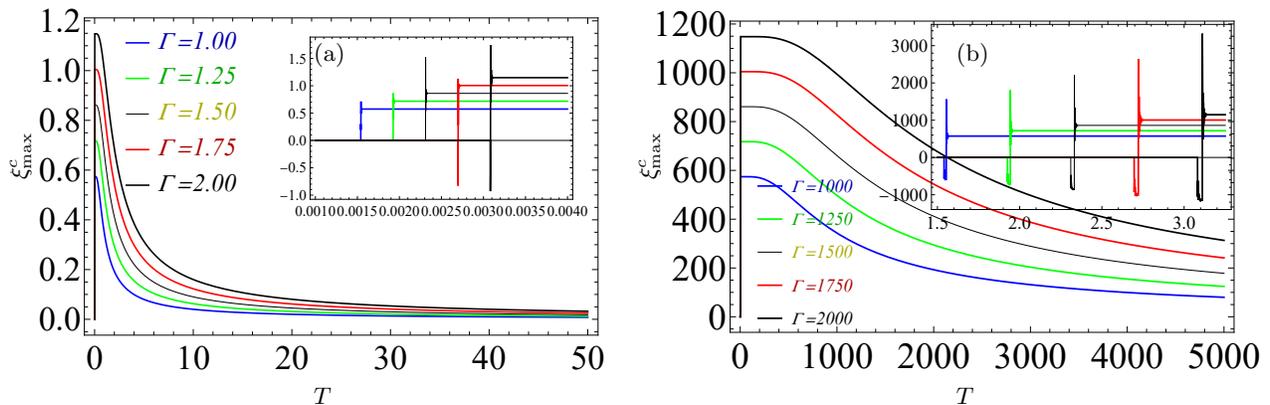
\begin{figure*}[t]
  \centering  
  \includegraphics[width=0.42\textwidth]{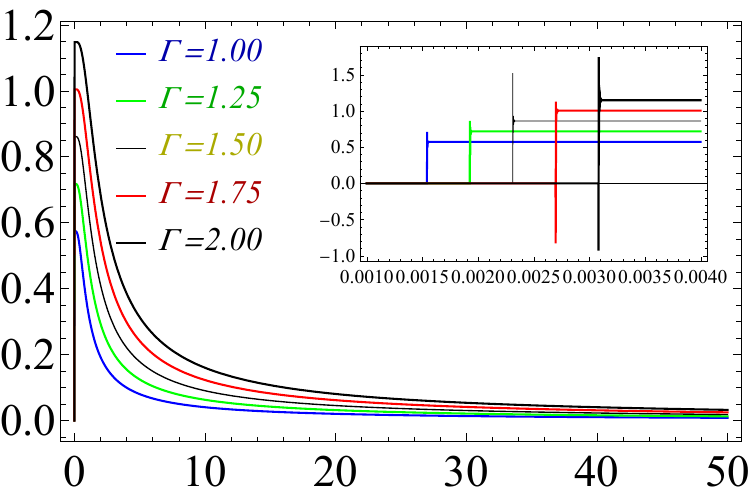}
  \put(-225,75){\rotatebox{90}{$\xi_{\max}^{c}$}}
  \put(-110,120){(a)} 
  \put(-110,-10){$T$} 
  \put(-140,40){$$} 
  \qquad  
  \includegraphics[width=0.45\textwidth]{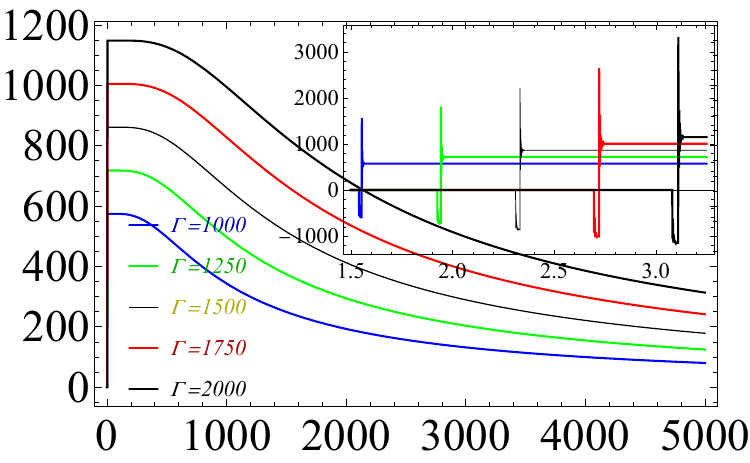}
  \put(-235,75){\rotatebox{90}{$\xi_{\max}^{c}$}}
  \put(-115,120){(b)} 
  \put(-115,-10){$T$} 
  \put(-140,33){$$} 
  \qquad
\caption{Evolution of maximum value of ergotropy in time, $\xi_{\max}^{c}$, as a function of $T$ for different fixed values of $\Gamma=1$ (blue), $\Gamma=1.25$ (green), $\Gamma=1.5$ (yellow), $\Gamma=1.75$ (red), and $\Gamma=2$ (black) at $\gamma=0.5$, $B=J=\delta=0$, $k=7\pi/8$, and $\Omega=1$. Here (a) is the AFM and (b) is FM case.}
\label{F4}
\end{figure*}

At $J = 0$, often referred to as the parallel charging situation in QB, each subsystem such as spins in this case, is charged independently. The maximum value of ergotropy, $\xi_{\max}^{c}$, is observed to be zero exclusively when $B = 0$ (represented by the black curve in Fig. \ref{F1}). This is because, at $B = 0$, the system lacks both complete and partial non-degenerate energy states. However, for $B > 0$, $\xi_{\max}^{c}$ increases as the external magnetic field disrupts the energy level degeneracy, with higher values of $B$ inducing a more significant increase in $\xi_{\max}^{c}$. 
In the collective charging scenario, where $J > 0$ (AFM case), the maximum ergotropy, $\xi_{\max}^{c}$, initially increases with $J$ up to a specific threshold. Beyond this point, the system undergoes a sudden quenching behavior in $\xi_{\max}^{c}$, followed by an abrupt rise, where $\xi_{\max}^{c}$ nearly doubles and then continues to increase linearly with a steeper slope. 
This oscillatory behavior in $\xi_{\max}^{c}$ occurs within the range $3.8 \lesssim J \lesssim 5.2$ at a low temperature of $T = 0.01$. However, as the temperature rises to $T = 0.1$, the threshold shifts to a higher range, $40 \lesssim J \lesssim 41.5$. This indicates that as thermal fluctuations increase, the sudden transitions in ergotropy shift to larger values of $J$. Therefore, at lower temperatures, where quantum correlations are more robust, even a slight increase in $J$ can lead to significant gains in ergotropy. However, at higher temperatures, the enhanced thermal noise reduces quantum coherence, diminishing the system’s sensitivity to $J$ and requiring higher values of $J$ to reach similar ergotropic transitions. 
Our observation of the sudden quenching leading to a sudden change in the ergotropy is consistent with the previous studies in the many-body spin quantum systems \cite{Hoang2024, Defenu2024}.
Oscillations associated with the quenching are attributed to the small sample sizes, so-called finite-size effects, but it is clear that work and ergotropy quickly rise when a quench knocks the system well out of equilibrium.

In the FM case ($J < 0$), as shown in panels \ref{F1}(b) and \ref{F1}(d), $\xi_{\max}$ initially increases with increasingly negative values of $J$ up to a certain threshold. Beyond this threshold, the system undergoes a quenching behavior accompanied by an abrupt transition. Unlike the positive $J$ case, where $\xi_{\max}^{c}$ continues to increase after the transition, here the ergotropy sharply rises immediately after quenching, only to rapidly decrease again for more negative values of $J$. This minimum value reaches zero when $B = 0$, whereas for $B > 0$, the minimum value remains above zero but is still significantly reduced. 
The threshold for this transition occurs within the range $-4.2 \lesssim J \lesssim -3.2$ at $T = 0.01$, shifting to $-40.25 \lesssim J \lesssim -39$ at $T = 0.1$. This shift suggests that like the positive $J$ case, the transitions are pushed to larger negative values of $J$ with increasing thermal fluctuations. Moreover, we find that ergotropy is generally greater for $|J| > 0$, corresponding to the collective charging regime, compared to $J = 0$, which represents parallel charging.

To analyze how variations in the anisotropy parameter $\delta$ affect the maximum ergotropy in time $\xi_{\max}^{c}$ as a function of the coupling strength $J$, as shown in Fig.~\ref{F2}(a) for the AFM case and Fig.~\ref{F2}(b) for the FM case, where $B = \Gamma = \gamma = 0$ and $T = 0.1$, we report the following observations. For a larger anisotropy value, say $\delta = 0.5$ (blue curve), the sudden transition in $\xi_{\max}^{c}$, as previously discussed, occurs at a relatively lower value of $J$ compared to a smaller value of $\delta$, for example, $\delta = 0.1$ (black curve). From Figs.~\ref{F1} and \ref{F2}, we can infer that while a larger $\delta$ leads to these sharp transitions in ergotropy within a narrower range of $J$, increasing the external field $B$ results in quenching, followed by abrupt transitions in $\xi_{\max}^{c}$ for larger $J$ values in both AFM and FM cases. This shows that the impact of anisotropy $\delta$ is beneficial, whereas that of the Zeeman field strength $B$ is somehow detrimental.

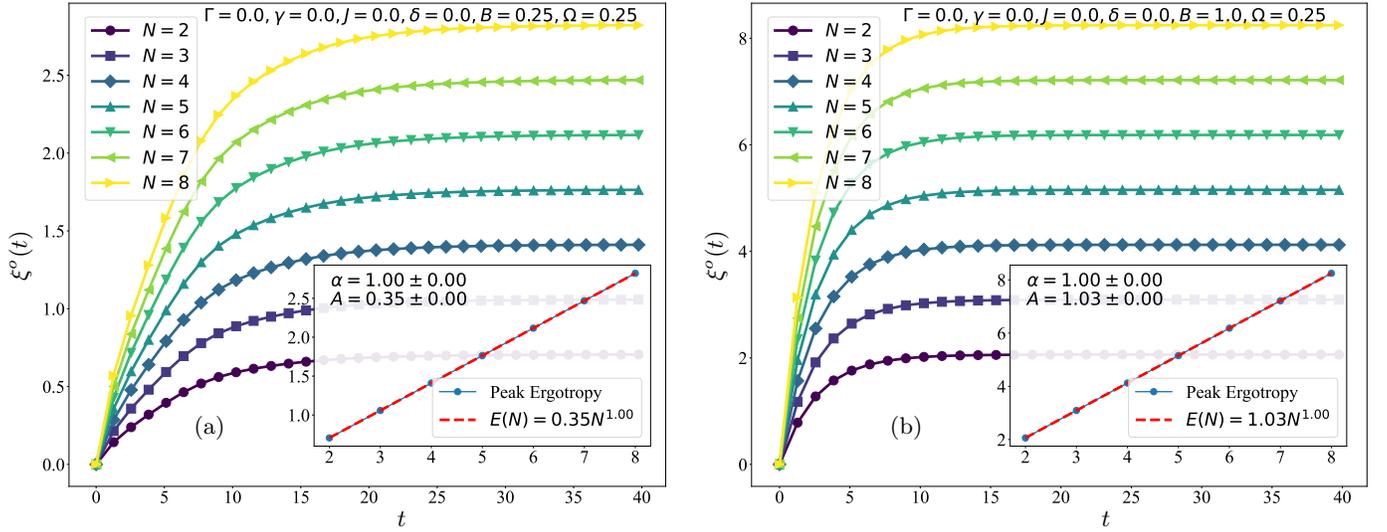
\begin{figure*}[t]
  \centering  
  \includegraphics[width=0.48\textwidth]{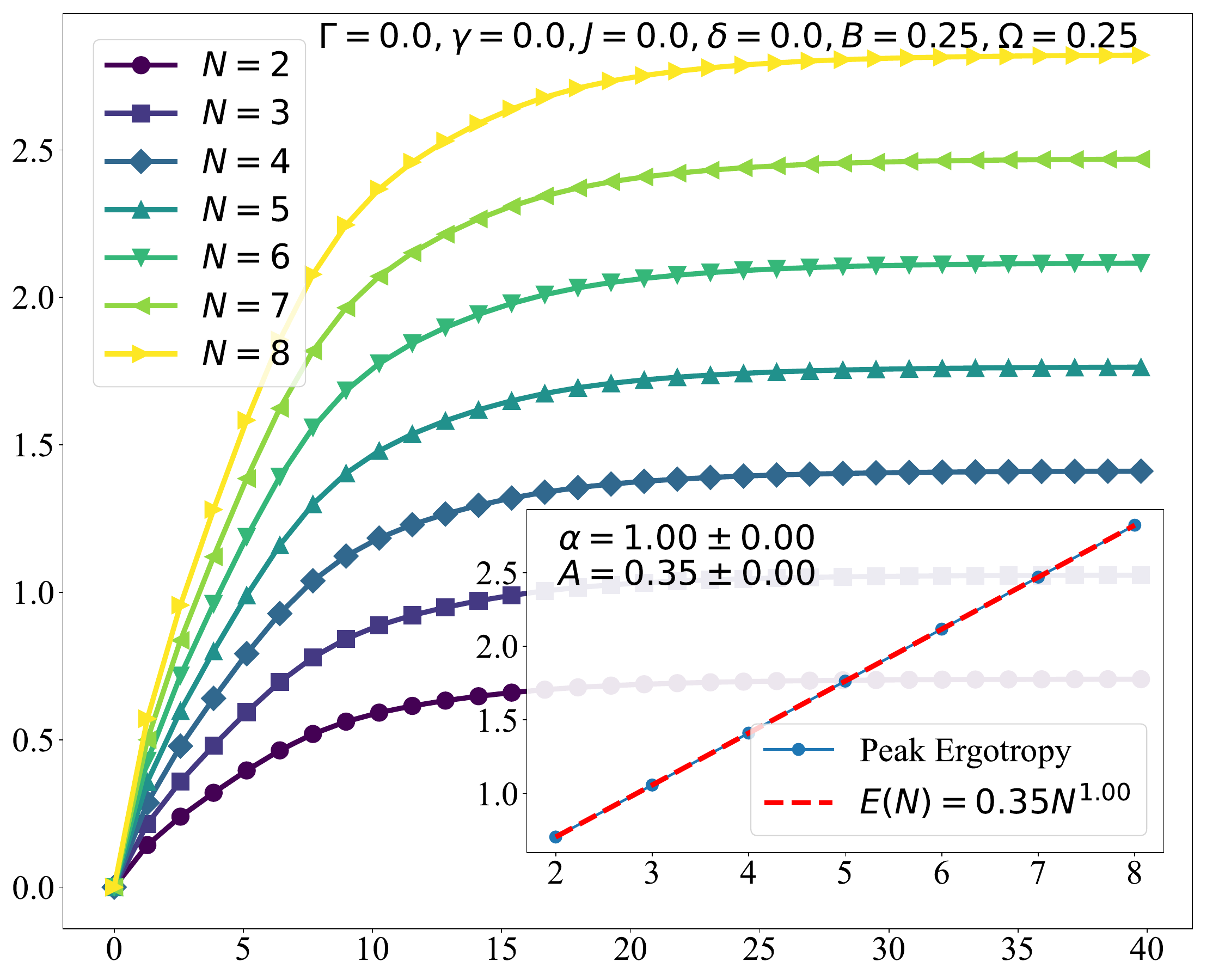}
  \put(-255,88){\rotatebox{90}{$\xi^{o}(t)$}}
  \put(-30,170){} 
  \put(-108,-5){$t$} 
  \put(-185,30){(a)} 
  \qquad  
  \includegraphics[width=0.48\textwidth]{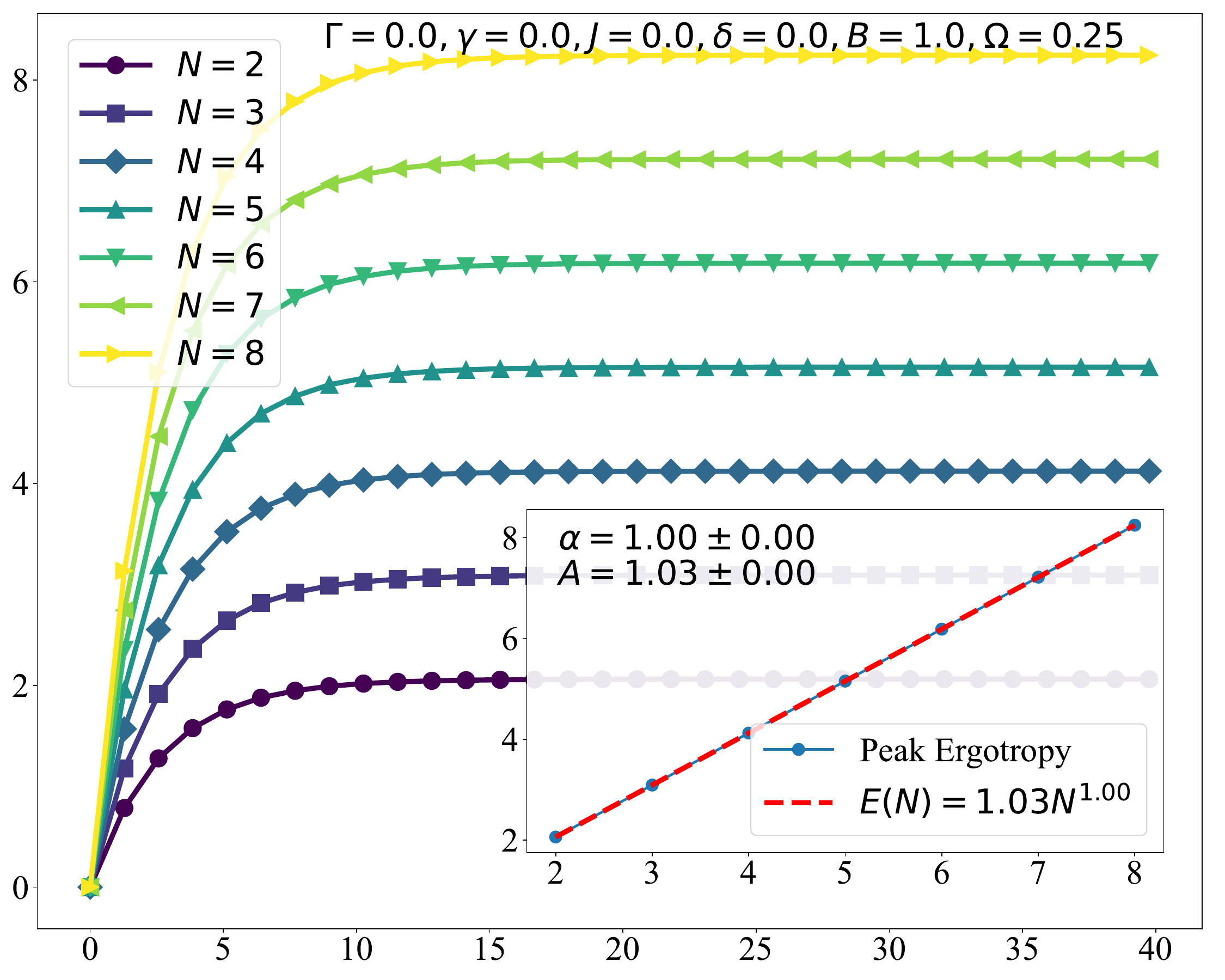}
  \put(-255,88){\rotatebox{90}{$\xi^{o}(t)$}}
  \put(-30,170){} 
  \put(-105,-5){$t$} 
  \put(-185,30){(b)} 
  \qquad
\caption{ The plots of $\xi^{o}(t)$ for different system size $N$ as a function of time $t$ for (a) $B = 0.25$, and (b) $B = 1$ at $\Gamma = \gamma = J = 0$, $T=0.1$ and $\Omega=0.25$. In the inset, blue dots represent the peak ergotropy $\xi_{\max}^{o}$ against each $N$ value, with the fitting function $E(N)$ plot shown as a thin red curve, illustrating the extensive scaling behavior.
}
\label{F5}
\end{figure*}

We now investigate how a change in $\Gamma$ affects $\xi_{\max}^{c}$ for different fixed values of $\gamma$ when $J=B=0$. Fig. \ref{F3} plots the variation of $\xi_{\max}^{c}$ with $\Gamma$ for different fixed values of $\gamma$ such as $\gamma=-1$ (blue), $\gamma=-0.5$ (green), $\gamma=0$ (yellow), $\gamma=0.5$ (red), and $\gamma=1$ (black). We observe that for negative values of $\gamma$, such as $\gamma=-1$ (DM interaction) and $\gamma=-0.5$, the increase in the $\Gamma$ increases $\xi_{\max}^{c}$ without any quenching, followed by a sudden transition. However, for $1\geq \gamma\geq0$, one can see that after certain values of $\Gamma$, quenching followed by sudden transition is observed, where larger values of $\gamma$, i.e. $\gamma=1$ (KSEA interaction), show this behavior for smaller values of $\Gamma$, while $\gamma=0$ shows similar quenching followed by a sudden rise in $\xi_{\max}^{c}$ for larger values of $\Gamma$. 

We further invetigate the impact of temperature $T$ on $\xi_{\max}^{c}$ for different 
values of $\Gamma$ -- we set $\gamma = 0.5$ for convenience. In Fig. \ref{F4}(a), $\Gamma$ is varied from 1 to 2 in steps of 0.25, while in Fig. \ref{F4}(b), it increases by three orders of magnitude, 
to highlight its substantial effect. In general, increasing $T$ tends to negatively impact the increase in $\xi_{\max}^{c}$, although for lower values of $T$, $\xi_{\max}^{c}$ initially increases to a peak before dropping off, marked by sudden quenches followed by sharp transitions. By increasing $\Gamma$, this quenching and transition behavior can be shifted to significantly higher $T$ values, as shown in Fig. \ref{F4}(b). This implies that by tuning $\Gamma$, we can achieve a substantial increase in maximum work output even at elevated temperatures. Therefore, we conclude that one can achieve a considerable improvement by operating the QB in this regime of sudden transitions to high ergotropy.

\begin{figure*}[t]
  \centering  
  \includegraphics[width=0.48\textwidth]{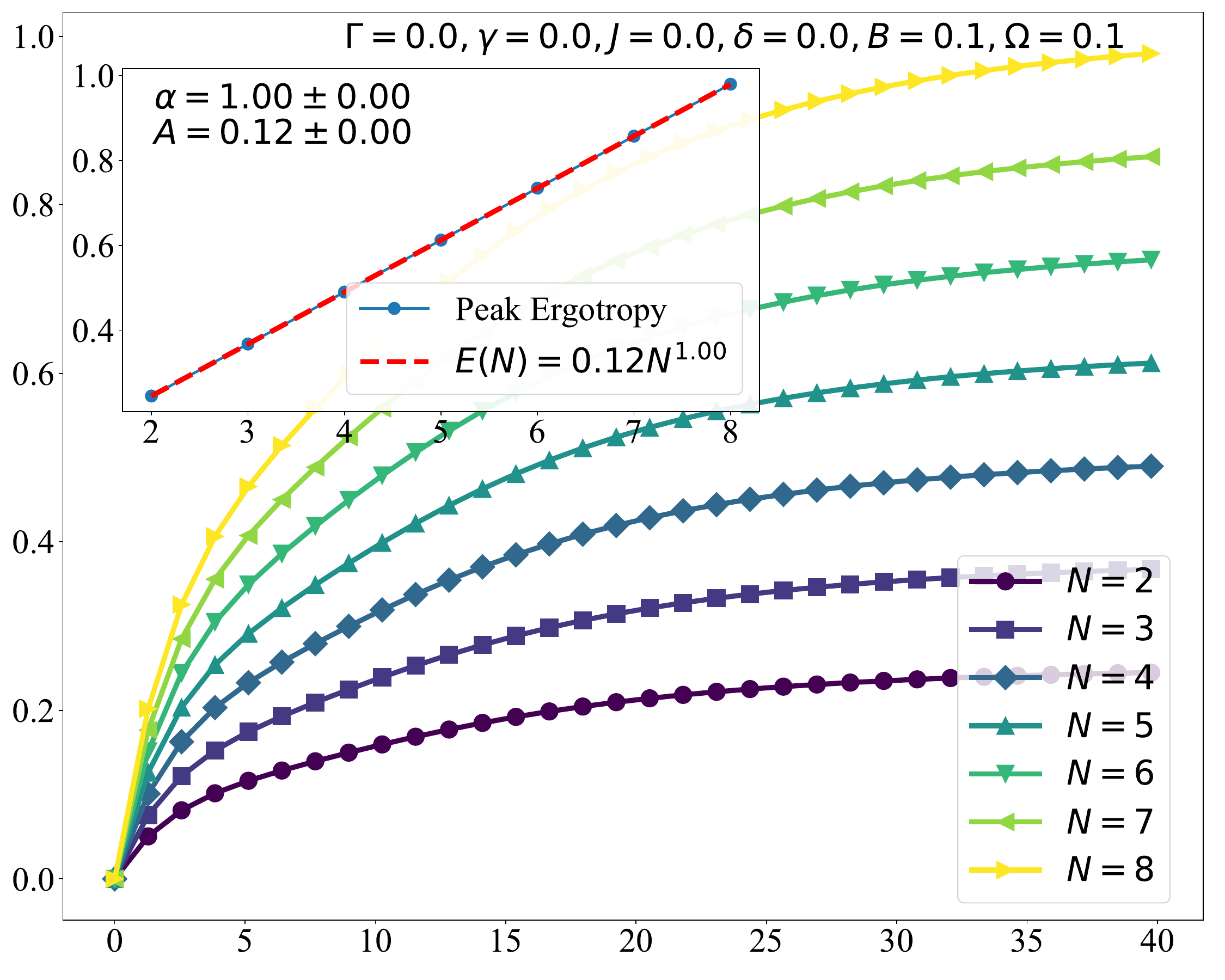}
  \put(-255,88){\rotatebox{90}{$\xi^{o}(t)$}}
  \put(-30,170){} 
  \put(-108,-5){$t$} 
  \put(-200,150){(a)} 
  \qquad  
  \includegraphics[width=0.48\textwidth]{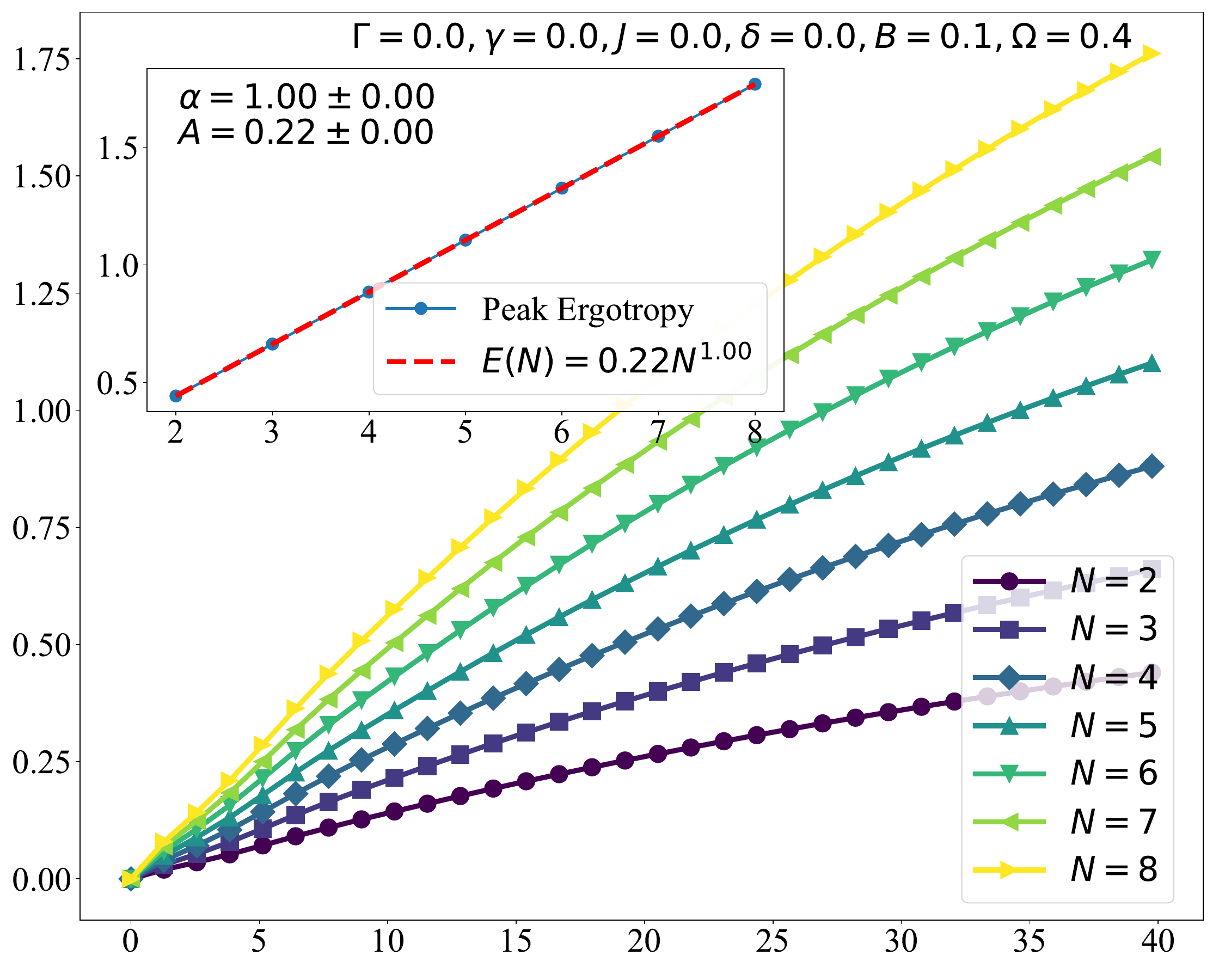}
  \put(-255,88){\rotatebox{90}{$\xi^{o}(t)$}}
  \put(-30,170){} 
  \put(-105,-5){$t$} 
  \put(-200,150){(b)} 
  \qquad
\caption{The plots of $\xi^{o}(t)$ for different system size $N$ as a function of time $t$ for (a) $\Omega = 0.1$, and (b) $\Omega = 0.4$  at $\Gamma = \gamma = J = 0$ and $B=0.1$. In the inset, blue dots represent the peak ergotropy $\xi_{\max}^{o}$ against each $N$ value, with the fitting function $E(N)$ plot shown as a thin red curve, illustrating the extensive scaling behavior.}
\label{F6}
\end{figure*}
\begin{figure*}[t]
  \centering  
  \includegraphics[width=0.48\textwidth]{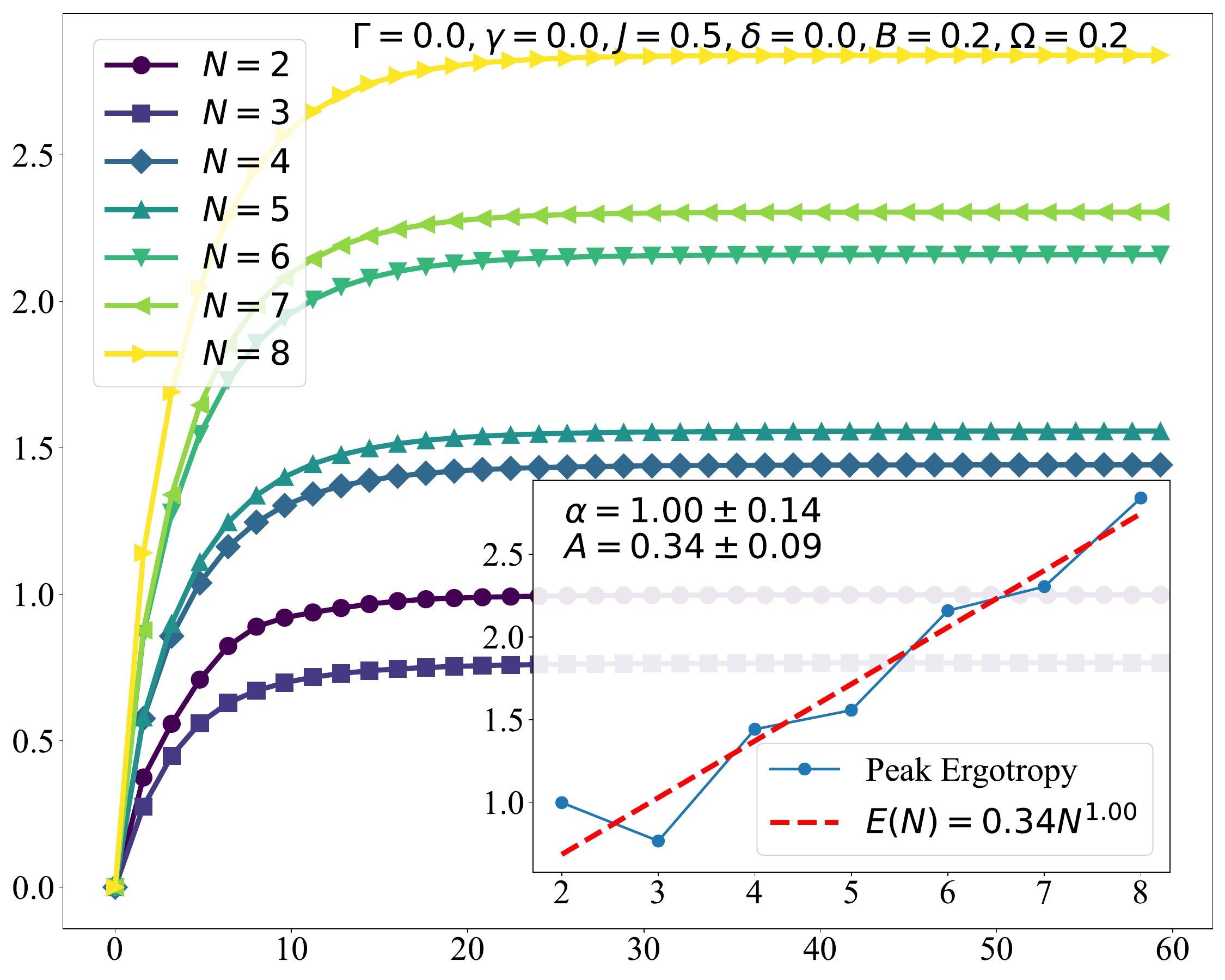}
  \put(-255,88){\rotatebox{90}{$\xi^{o}(t)$}}
  \put(-30,170){} 
  \put(-108,-5){$t$} 
  \put(-40,170){(a)} 
  \qquad  
  \includegraphics[width=0.48\textwidth]{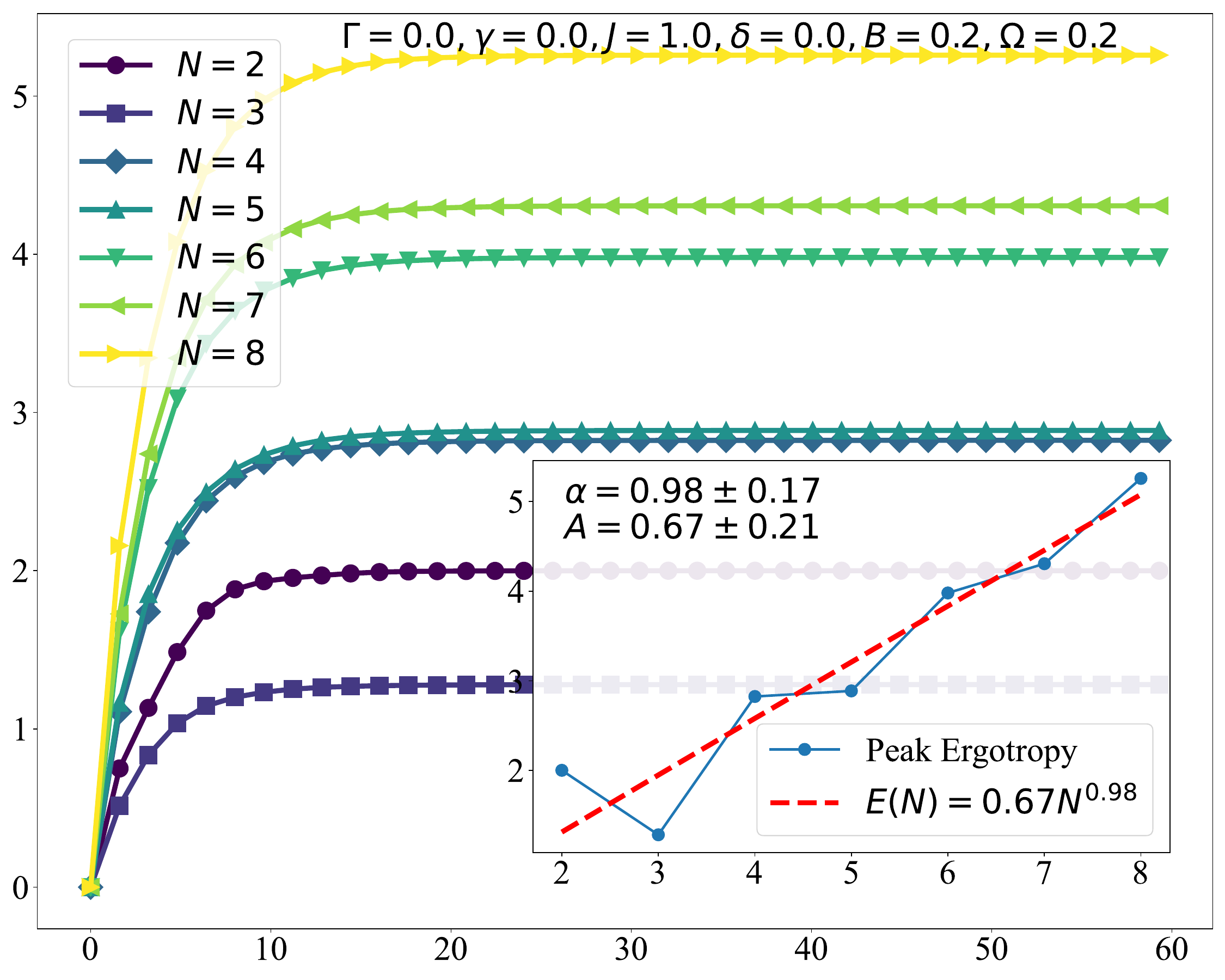}
  \put(-255,88){\rotatebox{90}{$\xi^{o}(t)$}}
  \put(-30,170){} 
  \put(-105,-5){$t$} 
  \put(-40,170){(b)} 
  \qquad
\caption{The plots of $\xi^{o}(t)$ for different system size $N$ as a function of time $t$ for $J=0.5$ (a) and $J = 1$ (b) at $\Gamma = \gamma=\delta= 0$, $T=0.1$ and $B=0.2$. In the inset, blue dots represent the peak ergotropy $\xi_{\max}^{o}$ against each $N$ value, with the fitting function $E(N)$ plot shown as a thin red curve, illustrating the extensive scaling behavior. The error in $\alpha$ is due to not having a perfect fit since the data is non-linear which gives residual least square errors and uncertainties.}
\label{F7}
\end{figure*}

\begin{figure*}[t]
  \centering  
  \includegraphics[width=0.48\textwidth]{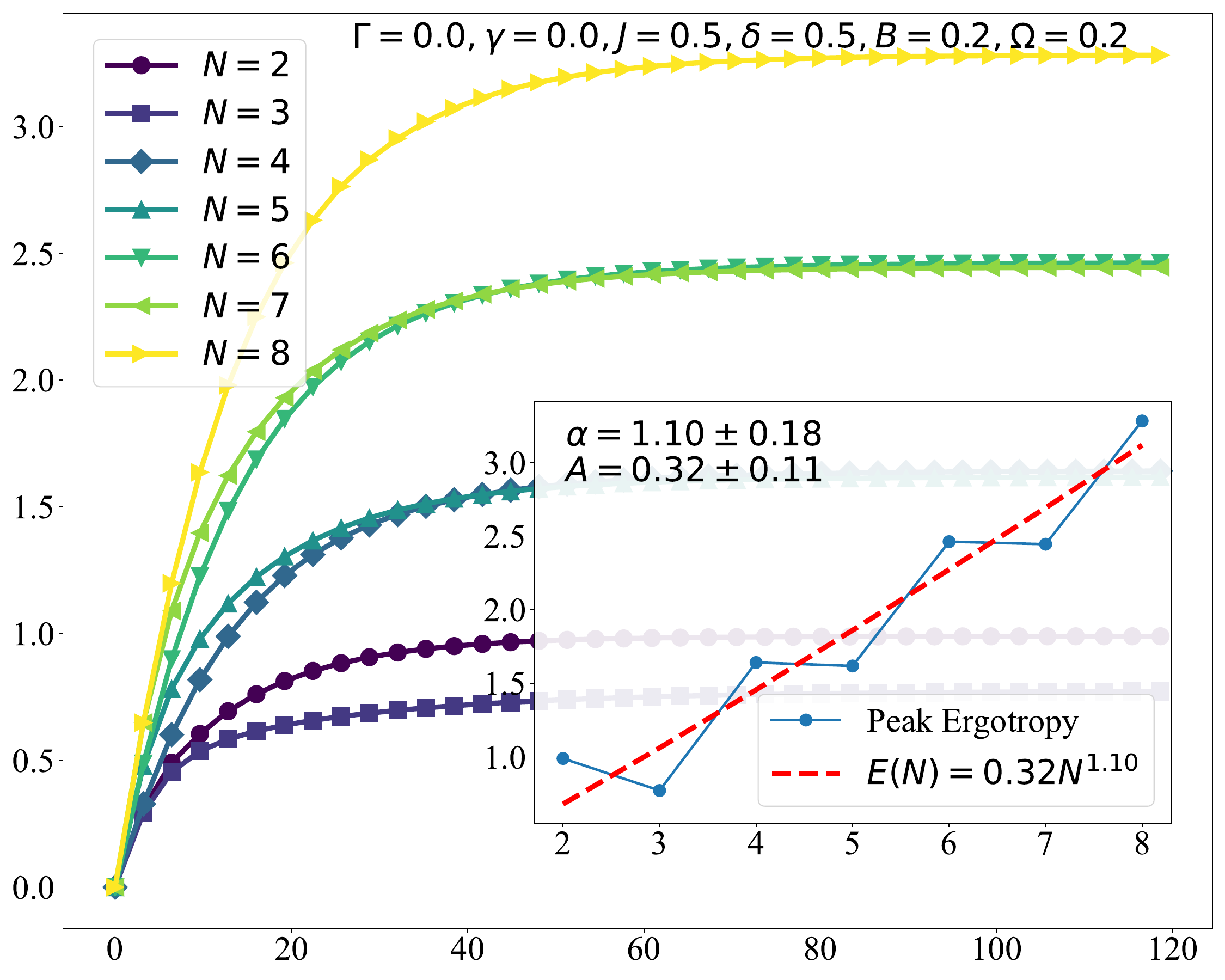}
  \put(-255,88){\rotatebox{90}{$\xi^{o}(t)$}}
  \put(-30,170){} 
  \put(-108,-5){$t$} 
  \put(-40,160){(a)} 
  \qquad  
  \includegraphics[width=0.48\textwidth]{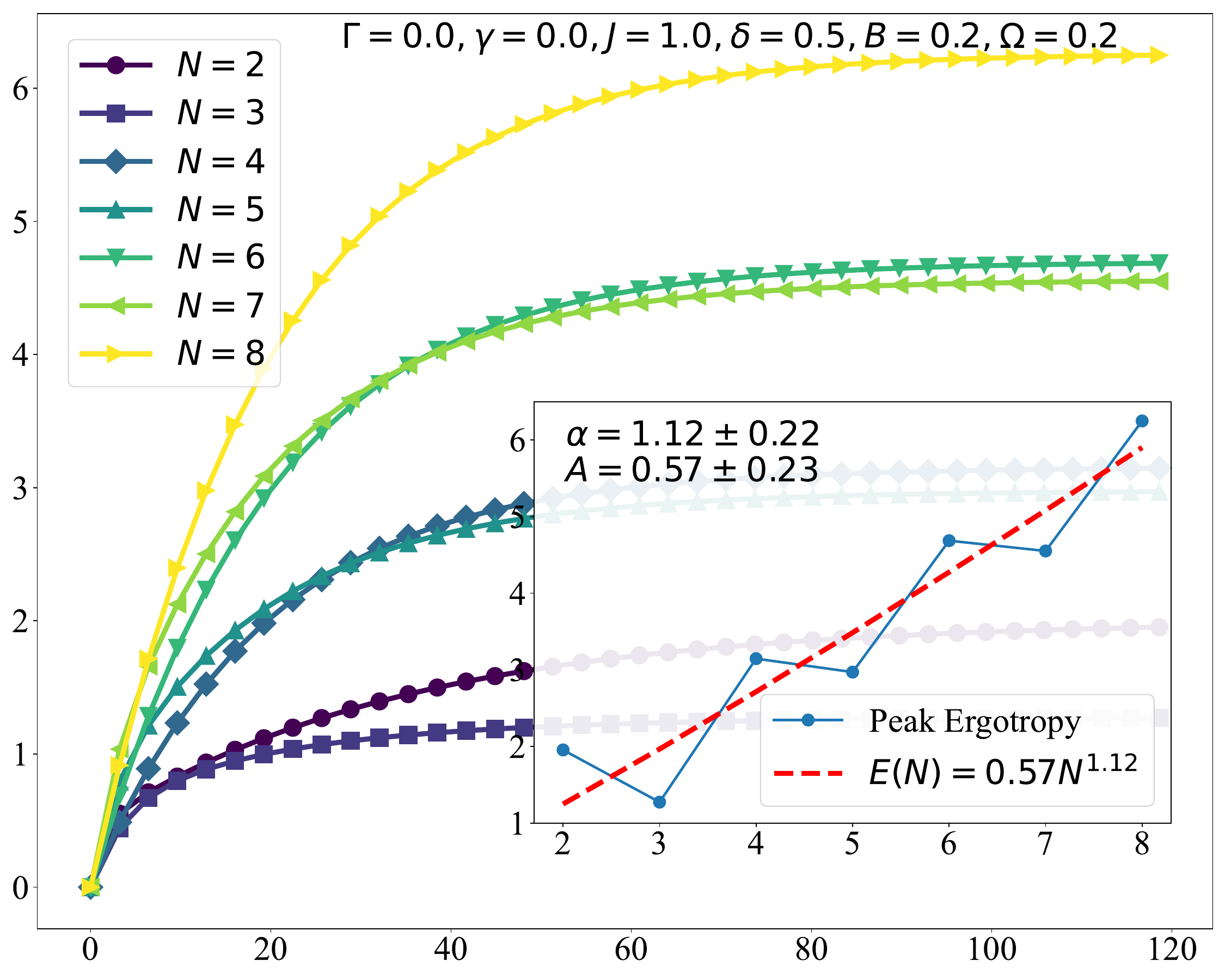}
  \put(-255,88){\rotatebox{90}{$\xi^{o}(t)$}}
  \put(-30,170){} 
  \put(-105,-5){$t$} 
  \put(-40,160){(b)} 
  \qquad
\caption{The plots of $\xi^{o}(t)$ for different system size $N$ as a function of time $t$ for $J=0.5$ (a) and $J = 1$ (b) at $\delta=0.5$, $\Gamma = \gamma= 0$, $T=0.1$ and $B=\Omega=0.2$. In the inset, blue dots represent the peak ergotropy $\xi_{\max}^{o}$ against each $N$ value, with the fitting function $E(N)$ plot shown as a thin red curve, illustrating the extensive scaling behavior.}
\label{F8}
\end{figure*}
\begin{figure*}[t]
  \centering  
  \includegraphics[width=0.48\textwidth]{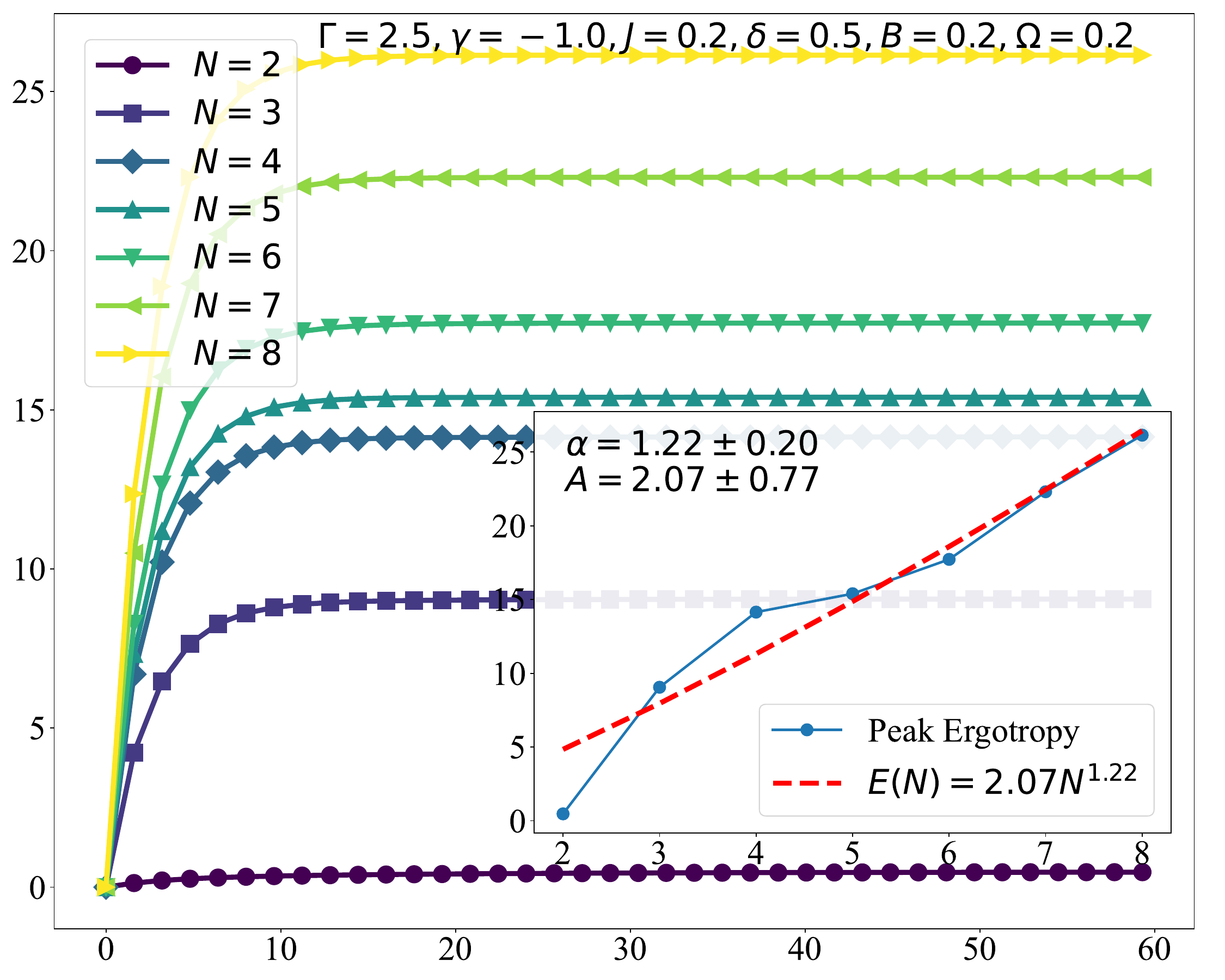}
  \put(-255,88){\rotatebox{90}{$\xi^{o}(t)$}}
  \put(-30,170){} 
  \put(-108,-5){$t$} 
  \put(-130,90){(a)} 
  \qquad    
  \includegraphics[width=0.48\textwidth]{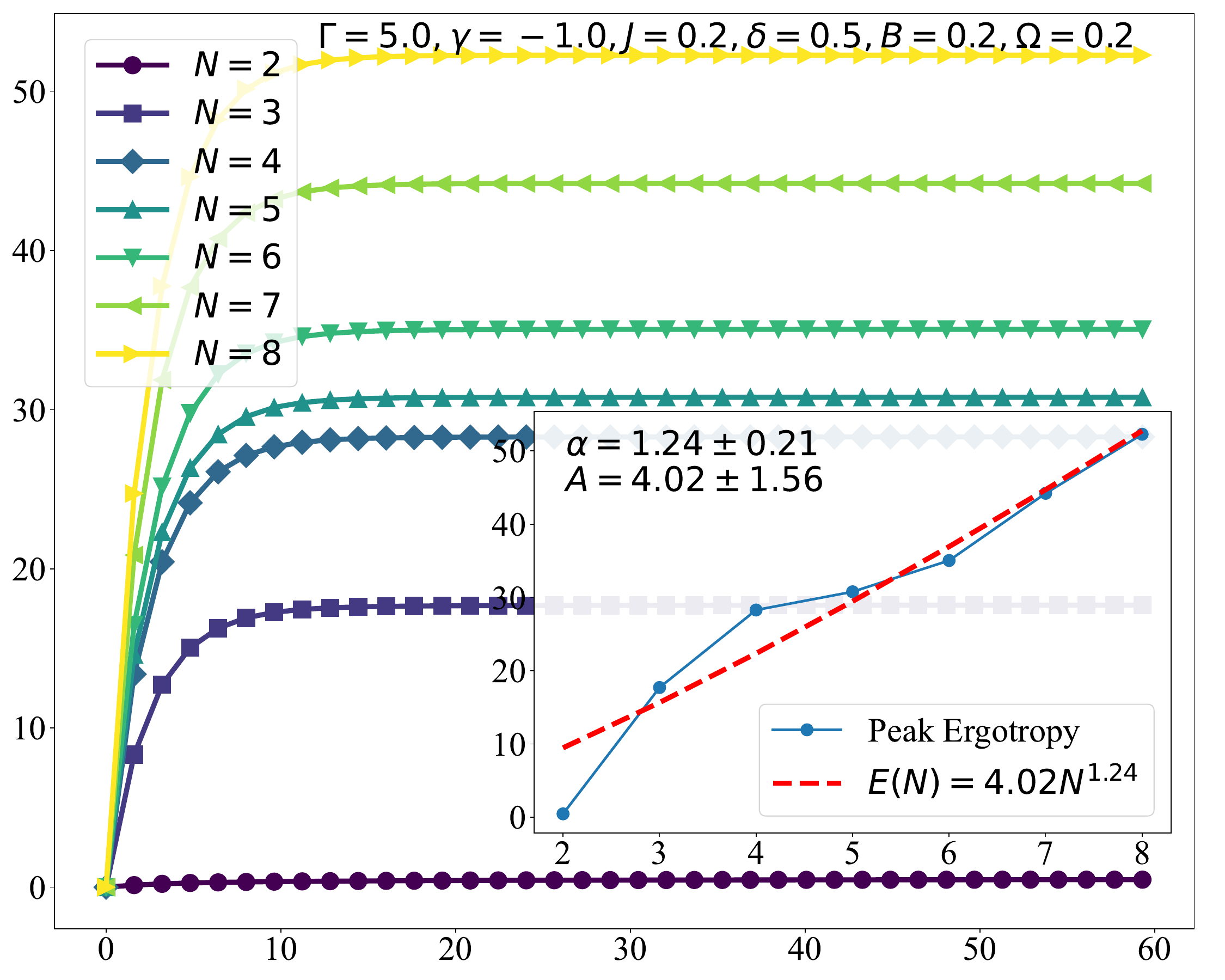}
  \put(-255,88){\rotatebox{90}{$\xi^{o}(t)$}}
  \put(-30,170){} 
  \put(-105,-5){$t$} 
  \put(-130,90){(b)} 
  \caption{The plots of $\xi^{o}(t)$ for different system sizes $N$ as a function of time $t$ for $\Gamma=2.5$ (a) and $\Gamma = 5$ (b) at $\gamma=-1$, $J=0.2$, $\delta= 0.5$, $T=0.1$ and $B=\Omega=0.2$. In the inset, blue dots represent the peak ergotropy $\xi_{\max}^{o}$ against each $N$ value, with the fitting function $E(N)$ plot shown as a thin red curve, illustrating the extensive scaling behavior.}
  \label{F9}
\end{figure*}

\subsection{Open QB scenario}
We now numerically analyze the ergotropy for spin chains with $N = 2$ to $N = 8$, increasing by steps of 1. Due to the exponential rise in computational complexity with $N$ numerous system, noise, and QB parameters, our computational resources limit us to $N = 8$. We focus on evaluating the performance of the open QB under Pauli $X$ noise using Eq. \eqref{eq: pauli_lind}, where we set $g=0.2$ throughout the analysis.

We evaluate ergotropy and its dynamics for $2 \leq N \leq 8$ as shown in Fig.~\ref{F5} 
when no spin-spin interaction exists ($J = \Gamma= \gamma = 0$). The charging field strength is set to $\Omega = 0.25$, and we examine how changing the Zeeman splitting field $B$ impacts the performance of the open QB. In \ref{F5}(a), we have set $B=0.25$, whereas for \ref{F5}(b),  $B=1$. We found that when $t=0$, the ergotropy is zero since the initial state of the QB is considered to be a Gibbs thermal state, which is a completely passive state, and thus we cannot extract any work out of it. However, with $t>0$, the ergotropy starts to increase until a steady state for generally large values of $t$. We find that since $J=0$ and $\Gamma=0$, it is a completely non-interacting spin QB scenario, this kind of situation in QB is called a parallel charging scenario. If we compare \ref{F5}(a) $B=0.25$ with \ref{F5}(b), $B=1$ we find that in this parallel charging QB scenario, an increase in Zeeman splitting not only increases the peak ergotropy of open QB for any $N$ but also increases the charging rate from Gibbs thermal state to steady state, where ergotropy reaches a maximum. 

After computing the ergotropy over time for different system sizes $N$, the peak ergotropy is extracted for each size. The objective is to fit these peak values to a fitting function or power law:
\begin{equation}
E(N) = A N^{\alpha}.    
\end{equation}
In this context,  $E(N)$ represents the fitting function for the peak ergotropy at a given system size $N$ over time, where $A$ is a constant and $\alpha$ is the scaling exponent. The scaling exponent $\alpha$ quantifies how the peak ergotropy varies with the number of spins $N$. A curve fitting procedure, optimized for maximum goodness of fit, is used to estimate $A$ and $\alpha$ from the peak ergotropy values in different $N$. Furthermore, uncertainties in $A$ and $\alpha$ are quantified using the covariance matrix of the fit, providing confidence intervals for the parameters. However, due to the model's complexity and the limitation to 8 spins, these uncertainties may be relatively inflated. While the approximate values of $\alpha$ and $A$ are discussed, their precise estimation would require simulations with larger $N$, which is beyond the scope of this study due to resource limitations. In the subsequent analysis of superextensive scaling, the bare $\alpha$ and $A$ values are considered without emphasizing the uncertainties in their values, acknowledging these limitations.

Each inset in Figs. \ref{F5}(a) and \ref{F5}(b) displays two plots: the first depicts the peak ergotropy over time for different system sizes $N$, represented by thick blue dots. The scaling exponent is determined to be $\alpha = 1$, signifying linear or extensive scaling, where the peak ergotropy increases linearly with QB size $N$. Furthermore, as the parameter $B$ increases fourfold from $B = 0.25$ to $B = 1$, the rate of charging, approximated by the slope $A$, also rises from $A = 0.25$ in Fig. \ref{F5}(a) to $A = 1.03$ in Fig. \ref{F5}(b). Thus, we conclude that even in non-interacting spin-based QBs (where $J = \Gamma = 0$), an increase in Zeeman splitting enhances the achievable ergotropy. However, this advantage is purely classical, as $\alpha$ does not exceed 1, reflecting the absence of coherence between spins due to $J = \Gamma = 0$.

In the parallel charging scenario where $J = \Gamma = 0$ with
negligible Zeeman field, i.e., $B = 0.1$, we observe that increasing charging field strength $\Omega = 0.1$ as shown in Fig. \ref{F6}(a) to $\Omega = 0.4$ in Fig. \ref{F6}(b), results in $\alpha = 1$, even in the absence of spin-spin interaction. However, $A$ increases from 0.12, as shown in the inset of Fig. \ref{F6}(a), to 0.22 in Fig. \ref{F6}(b), indicating an increase in the charging rate. This confirms the absence of the quantum advantage as $\alpha$ does not exceed 1.

Now we consider the situation where $J \neq 0$ but the anisotropy in $J$ denoted by $\delta=0$. This corresponds to the $XX$ model where spin-spin couplings are isotropic. In QB analysis, this situation is often known as collective charging. For convenience and brevity, we consider the AFM case and study how the performance of the open QB is affected without the presence of $\Gamma$ interaction, therefore, we set $\Gamma = 0$.  We observe that increasing $J$ from $J = 0.5$ in Fig.~\ref{F7}(a) to $J = 1$ in Fig.~\ref{F7}(b) may or may not result in super-extensive scaling for given fixed values of other parameters, as the scaling exponent $\alpha$ remains below unity ($\alpha < 1$). Specifically, for $J = 0.5$, we obtain $\alpha = 1$ (linear scaling), while for $J = 1$, $\alpha$ decreases to $0.98$ (sub-linear scaling). This indicates that increasing $J$ under zero anisotropy alone for given fixed values of other parameters is not sufficient to achieve super-extensive scaling. Indeed, larger values of $J$ may lead to sub-extensive scaling, as evidenced in Fig.~\ref{F7}(b) with $\alpha = 0.98$ for given fixed values of parameters. We notice that although a higher $J$ enhances the charging rate, as reflected by an increase in $A$ from $0.34$ in Fig.~\ref{F7}(a) to $0.67$ in Fig.~\ref{F7}(b), this improvement in $A$ does not correspond to super-extensive scaling. 

In contrast to Fig. \ref{F7}, where we set $\delta = 0$, we now consider $\delta = 0.5$. In this case, the scaling exponent becomes $\alpha = 1.1$ as shown in Fig. \ref{F8}(a) for $J=0.5$. This value can be further enhanced by increasing the spin-spin coupling to $J = 1$, as shown in Fig. \ref{F8}(b), where $\alpha = 1.14$. These results indicate that to achieve and strengthen super-extensive scaling, it is necessary not only to increase the spin-spin coupling but also to introduce anisotropy in spin-spin coupling in the system by setting $\delta = 0.5$. This represents maximum anisotropy in the spin-spin interaction in the $XY$ plane because the larger value of $J$ will result in faster charging rate, whereas high anisotropy, i.e. $\delta=0.5$, will assist in super-extensive scaling. This is in contrast to $\delta = 0$, which corresponds to isotropic spin-spin coupling in the $XY$ plane. 

Finally, we investigate the impact of the interaction parameter $\Gamma$ on the system. The ergotropy as a function of time is plotted in Fig. \ref{F9}, with fixed values of $J = 0.2$, $ B = 0.2 $, and $\Omega = 0.2$, while setting $ \gamma = -1$ (DM interaction). In Fig. \ref{F9}(a), where $\Gamma = 2.5$, the extensive scaling exponent $\alpha$ is found to be 1.24. In Fig. \ref{F9}(b), for $ \Gamma = 5.0$, $\alpha$ increases further to 1.24, reflecting a 24\% increase in ergotropy. These results demonstrate that the presence of a non-zero $\Gamma$ enables the system to achieve larger super-extensive scaling, thereby enhancing the charging performance of the QB when operating in the $-1 < \gamma \leq 1$ regime, except at $\gamma=0$. Therefore, we conclude that the quantum advantage in the form of superextensive scaling can be achieved by a non-zero value of  $\gamma$ with $\Gamma \neq0$.

\section{Conclusion}\label{sec4}

In this paper, we explored the potential of the 1D spin$-1/2$ Heisenberg $XY-\Gamma(\gamma)$ chain as a working medium for a QB also known as Kitaev quantum battery (KQB) and analyzed its performance in closed and open system scenarios. The initial state of QB is considered to be a Gibbs state and the QB is charged through Pauli-$X$ based charging Hamiltonian.

For the closed QB scenario, we diagonalized the system and picked two out of N-spin 1D spin 1/2 Kitaev lattice. We analytically evaluated ergotropy and analyzed it by independently varying different model parameters such as spin-spin coupling, anisotropy in spin-spin coupling, Zeeman splitting, charging field strength, $\Gamma$ interaction, and the temperature of QB, which significantly affect the peak ergotropy.
We showed out-of-equilibrium characteristics linked to quenches in the peak ergotropy with spin and Kitaev couplings as control parameters---which are in agreement with previous investigations related to quenched-induced quantum phase transitions.
In the AFM case, we observed that increasing the spin-spin coupling generally led to an increase in peak ergotropy over time, up to a certain threshold. Beyond this point, a sudden quenching type effect is observed which is followed by a sharp rise in the peak ergotropy. In contrast, in the FM case, increasing the spin-spin coupling also increases the peak ergotropy up to a threshold. However, after quenching, instead of an increase, the peak ergotropy decreases to a minimum. We find that these threshold values can be shifted for smaller or larger values of spin-spin coupling by varying the Zeeman splitting field and temperature. Furthermore, we find that the temperature negatively impacted the ergotropy. The non-zero $\Gamma$ interaction on peak ergotropy has a significantly positive impact on maximizing peak ergotropy. 

In the open QB case, we focused on the effects of Pauli-$X$ noise on their performance using spin chains ranging from two to eight spins. 
Initially, we observed zero ergotropy 
at $t$ = 0, due to the system being in a Gibbs thermal state. 
Over time, ergotropy increased until it attained its maximum value in a steady state.
We investigated two charging scenarios: parallel charging, characterized by non-interacting spins where all types of spin-spin coupling are zero, and collective charging, where spin interactions are present. In the parallel charging scenario, we found that increasing the Zeeman splitting field enhances both peak ergotropy and charging rates for all system sizes. However, this advantage remains classical as the scaling exponent does not exceed one, indicating a lack of coherence among spins. In contrast, during collective charging with spin-spin coupling being present, we discovered that merely increasing the spin-spin coupling does not necessarily lead to super-extensive scaling unless anisotropy in spin coupling is also introduced for given constant values of other parameters. Our findings suggest that achieving a quantum advantage in open QBs requires both anisotropic spin coupling and non-zero $\Gamma$ coupling. When these conditions are met, we observe improved charging performance and super-extensive scaling of ergotropy, demonstrating that quantum effects can significantly enhance the performance of QB. We found that the performance metric of the QB relates directly to the parametric optimization. Therefore, to achieve optimal objective functions, a rigorous analysis of QB parameters is necessary. Future research will explore the application of quantum optimal control theory to enhance the super-extensive scaling exponent beyond $\alpha=1.24$ in these KQBs. Additionally, we will investigate whether employing an entangled charging operator, as opposed to a local charging operator, offers any advantages in terms of performance and scalability.

\appendix

\section{Diagonalization of QB Hamiltonian}
\label{appendixA}
The Hamiltonian, $\mathcal{H}_\mathcal{QB}$, can be analytically solved using a Jordan-Wigner transformation:

\begin{equation}
    \sigma^+ = \exp\left[i\pi \sum_{m<n} c_m^\dagger c_m\right] c_n,
\end{equation}

\begin{equation}
    \sigma^- = 2c_n^\dagger c_{n-1},
\end{equation}
followed by a Fourier transformation:

\begin{equation}
    c_n = \frac{1}{\sqrt{N}} \sum_k \exp(ikn)c_k,
\end{equation}
where the possible values of $k$ should be given for a fixed value of $N$. Finally, Bogoliubov transformations are applied:

\begin{equation}
    c_k = \cos(\Phi_k)\eta_k - \sin(\Phi_k)e^{i\theta_k}\eta_{-k}, \tag{7a}
\end{equation}

\begin{equation}
    c_k^\dagger = \cos(\Phi_k)\eta_{-k} + \sin(\Phi_k)e^{-i\theta_k}\eta_k. \tag{7b}
\end{equation}
Therefore, we have
\begin{equation}
\begin{aligned}
&\left(\frac{1+\delta}{2}\right)\sigma_n^x \sigma_{n+1}^x 
+ \left(\frac{1-\delta}{2}\right)\sigma_n^y \sigma_{n+1}^y \\
&\quad = c_n^\dagger c_{n+1} + c_{n+1}^\dagger c_n 
+ \delta \left(c_n^\dagger c_{n+1}^\dagger + c_{n+1} c_n \right)
\end{aligned}
\label{eq16}
\end{equation}
and
\begin{equation}
\begin{aligned}
\sigma_n^x \sigma_{n+1}^y + \gamma \sigma_n^y \sigma_{n+1}^x 
&= -i \big[(\gamma + 1) \left(c_n^\dagger c_{n+1}^\dagger - c_{n+1} c_n \right) \\
&\quad + (\gamma - 1) \left(c_n^\dagger c_{n+1} - c_{n+1}^\dagger c_n \right)\big].
\end{aligned}
\label{a17}
\end{equation}

Fourier transformation $c_n = (1/\sqrt{N}) \sum_k e^{ikn}c_k$ on \eqref{eq16} and \eqref{a17} gives

\begin{equation}
\begin{aligned}
\mathcal{H}_{XY} + \mathcal{H}_{\mathcal{F}} 
&= \sum_{k > 0} \Big[\mathcal{A}_k\left(c_k^\dagger c_k + c_{-k}^\dagger c_{-k}\right) \\
&\quad + i\mathcal{B}_k\left(c_k^\dagger c_{-k}^\dagger + c_k c_{-k}\right) - 2B\Big],
\end{aligned}
\end{equation}
wherein we have substituted
\begin{equation}
\mathcal{A}_k = 2[J\cos(k) + B]
\end{equation}
and
\begin{equation}
\mathcal{B}_k = 2J\delta\sin(k).
\end{equation}

Likewise, the Fourier transformation  on \eqref{a17} gives

\begin{equation}
\begin{aligned}
\mathcal{H}_{\Gamma} 
&= 2\Gamma \sum_{k > 0} \sin(k) \Big[(\gamma + 1)(c_k^\dagger c_{-k}^\dagger - c_k c_{-k}) \\
&\quad + (\gamma - 1)(c_k^\dagger c_k - c_{-k}^\dagger c_{-k})\Big] \\
&= \sum_{k > 0} \Big[P_k(c_k^\dagger c_k - c_{-k}^\dagger c_{-k}) 
+ Q_k(c_k^\dagger c_{-k}^\dagger - c_k c_{-k})\Big],
\end{aligned}
\end{equation}
where we have considered 

\begin{equation}
P_k = 2\Gamma (\gamma - 1) \sin(k)
\end{equation}
and
\begin{equation}
Q_k = 2\Gamma (\gamma + 1) \sin(k).
\end{equation}

\section*{Disclosures}
The authors declare that they have no known competing financial interests.

\section*{Data availability}
The datasets can be made available from the corresponding author upon reasonable request.

\bibliography{bibliography}
\end{document}